\pdfoutput=1

\documentclass[11pt,twoside,a4paper,cmspaper,final,collab]{cms-tdr}
\def\svnVersion{55a9bf2-D}\def\svnDate{2020/06/06}\def\cmsCernNoTag{CERN-EP-2019-259}\def\cmsCernDate{\today}\def\cmsMessage{"Published in the Journal of High Energy Physics as \href{http://dx.doi.org/10.1007/JHEP06(2020)018}{\doi{10.1007/JHEP06(2020)018}.}"}

\begin{document}\cmsNoteHeader{SMP-18-005}

\hyphenation{had-ron-i-za-tion}
\hyphenation{cal-or-i-me-ter}
\hyphenation{de-vices}

\newcommand{\sqrts}{\ensuremath{\sqrt{s}}}
\newcommand{\mZ}{\ensuremath{m_{\PZ}}}
\newcommand{\mW}{\ensuremath{m_{\PW}}}
\newcommand{\MCSANC}{\textsc{mcsanc}\xspace}
\newcommand{\LHAPDF}{\textsc{lhapdf}\xspace}
\newcommand{\statt}{(\text{num})}

\providecommand{\cmsTable}[1]{\resizebox{\textwidth}{!}{#1}}
\newlength\cmsTabSkip\setlength{\cmsTabSkip}{1ex}

\hyphenation{ap-prox-i-mate-ly}
\hyphenation{HERAPDF}

\cmsNoteHeader{SMP-18-005}
\title{Determination of the strong coupling constant \texorpdfstring{$\alpS(m_{\PZ})$}{alphas(mZ)} from measurements
of inclusive \texorpdfstring{\PWpm}{W+-} and \texorpdfstring{$\PZ$}{Z} boson production cross sections in proton-proton collisions
at \texorpdfstring{$\sqrt{s} = 7$ and $8\TeV$}{sqrt(7) = 7 and 8 TeV}}

\date{\today}

\abstract{
Twelve measurements of inclusive cross sections of \PWpm and \PZ boson production, performed in proton-proton collisions at
centre-of-mass energies of 7 and 8\TeV, are compared with perturbative quantum chromodynamics  calculations at
next-to-next-to-leading order (NNLO) accuracy obtained with the CT14, HERAPDF2.0, MMHT14, and NNPDF3.0 parton distribution functions (PDFs).
Data and theory agree well for all PDF sets, taking into account the experimental and theoretical uncertainties.
A novel procedure is employed to extract the strong coupling constant at the \PZ pole mass from a detailed
comparison of all the experimental fiducial cross sections to the corresponding NNLO theoretical predictions, yielding
$\alpS(m_{\PZ}) = 0.1163^{+0.0024}_{-0.0031}$ (CT14), $0.1072^{+0.0043}_{-0.0040}$ (HERAPDF2.0), $0.1186\pm0.0025$ (MMHT14),
and $0.1147\pm 0.0023$ (NNPDF3.0). Using the results obtained with the CT14 and MMHT14 PDFs, which yield the most
robust and stable $\alpha_{S}(m_{\PZ})$ extractions, a value $\alpha_{S}(m_{\PZ}) = 0.1175^{+0.0025}_{-0.0028}$ is determined.
}

\hypersetup{
 pdfauthor={CMS Collaboration},
 pdftitle={Determination of the strong coupling constant alphas(mZ) from measurements of inclusive W and Z boson production cross sections
in proton-proton collisions at sqrt(s) = 7 and 8 TeV},
 pdfsubject={CMS},
 pdfkeywords={CMS, physics, W boson, Z boson, alphas}}

\maketitle

\section{Introduction}
\label{sec:intro}

In the chiral limit of zero quark masses, the $\alpS$ coupling is the only free parameter of quantum chromodynamics (QCD),
the theory of the strong interaction between quarks and gluons. Because of its logarithmic decrease with energy (asymptotic freedom),
$\alpS$ is commonly given at a reference scale, often taken at the \PZ pole mass. Its current value, $\alpS(m_{\PZ}) = 0.1181 \pm 0.0011$,
is known with a $\pm$0.9\% uncertainty, making it the least precisely known of all interaction couplings in nature~\cite{PDG}.
The precision of the strong coupling value plays an important role in all theoretical calculations of perturbative QCD (pQCD) processes involving
partons, and currently leads to 3--7\% uncertainties in key Higgs boson processes, such as the cross sections for gluon-gluon fusion ($\Pg\Pg\to \PH$)
and associated production with a top quark pair ($\ttbar\PH$),
as well as the $\PH\to\bbbar,\ccbar,\Pg\Pg$ partial decay widths~\cite{Dittmaier:2012vm}.
As one of the fundamental parameters of the standard model (SM), the uncertainties of the QCD coupling value also dominate the
propagated parametric uncertainties in the theoretical calculations of the top quark mass~\cite{Hoang:2017suc}, as well as of electroweak (EW)
precision observables~\cite{Gomez-Ceballos:2013zzn}.
Last but not least, $\alpS$ also impacts physics approaching the Planck scale, either through the EW vacuum stability~\cite{Buttazzo:2013uya},
or in searches of new coloured sectors that may modify its running towards the grand unification scale~\cite{Bourilkov:2015mua,Sannino:2015sel}.

The current $\alpS(m_{\PZ})$ world-average value is derived from a combination of six subclasses of (mostly) independent observables measured
at various energy scales, which are compared with pQCD calculations at next-to-next-to-leading order (NNLO), or beyond, accuracy~\cite{PDG}.
The only hadron collider observable so far that provides a constraint on $\alpS$ at this level of theoretical accuracy is the total $\ttbar$ cross
section~\cite{Chatrchyan:2013haa,Klijnsma:2017eqp,Sirunyan:2018goh}. One of the paths towards improvement of our knowledge of the QCD coupling is
the inclusion into the world average of new independent observables sensitive to $\alpS$ that are experimentally
and theoretically known with high precision~\cite{dEnterria:2015kmd,dEnterria:2019its}. Charged- and neutral-current Drell--Yan processes in their
leptonic decay modes, $\Pp\Pp \to \PWpm \to \ell^\pm \Pgn_{\ell}$ and $\Pp\Pp \to \PZ \to \ell^+\ell^-$ with $\ell^\pm=\Pepm,\PGmpm$,
are the most accurately known processes currently accessible in  proton-proton ($\Pp\Pp$) collisions at the CERN LHC.
Experimentally, the uncertainties in the inclusive \PV = \PWpm, \PZ production cross sections measured
by the CMS experiment are between 3 and 5\%; these are dominated by the integrated luminosity uncertainty,
whereas the statistical uncertainties are at the subpercent level (Table~\ref{tab:data})~\cite{CMS:2011aa,Chatrchyan:2014mua}.
On the theoretical side, the corresponding cross sections are known at NNLO pQCD accuracy~\cite{Hamberg:1990np},
with about 1--4\% parton distribution function (PDF), and 0.3--1.3\% scale
uncertainties~\cite{Boughezal:2016wmq}. Electroweak corrections, which lead to a few percent reduction of the
pure-pQCD \PWpm and \PZ boson production cross sections, are known at next-to-leading order (NLO) accuracy~\cite{Bondarenko:2013nu}.

Theoretical calculations~\cite{Anastasiou:2003ds} indicate that about one fourth of the total \PV
production cross sections at LHC energies come from partonic processes beyond the Born level,
and thereby depend on the QCD coupling value. By calculating the \PV
production cross sections at NNLO for varying $\alpS(m_{\PZ})$ values, and by comparing the theoretical
predictions to experimental data, one can therefore derive a value of the strong coupling constant at the \PZ pole,
independent of other current extractions~\cite{Poldaru:2019}.
By combining such a result with those derived from other methods, the overall uncertainty in the $\alpS(m_{\PZ})$ world average
can eventually be reduced.
The use of inclusive \PWpm, \PZ boson cross sections to extract the QCD coupling is presented here for the first time.
This method is similar to the one used to extract $\alpS$ from the inclusive $\ttbar$ cross sections at hadron
colliders~\cite{Chatrchyan:2013haa,Klijnsma:2017eqp,Sirunyan:2018goh}, except that the underlying physical process is quite different.
Whereas $\sigma(\ttbar)$ depends on $\alpS$ already at leading order (LO), albeit with $\approx$5\% theoretical
and experimental uncertainties, $\sigma(\PV)$
is more precisely known experimentally and theoretically, although at the Born level its underlying partonic processes are purely EW with a dependence on $\alpS$ that comes only through higher-order pQCD corrections (at LO, the $\sigma(\PV)$ cross sections also depend on $\alpS$ via the PDFs).

The paper is organised as follows. The experimental setup used in the twelve CMS original measurements is summarised in Section~\ref{sec:exp}.
In Section~\ref{sec:thsetup}, the theoretical tools used to perform the calculations are outlined. In Section~\ref{sec:cross_sections},
the experimental and theoretical cross sections
with associated uncertainties, are compared.
In Section~\ref{sec:4}, the method to extract $\alpS(m_{\PZ})$ from the data-to-theory comparison for each measurement
is described, as well as the approach to combine all $\alpS(m_{\PZ})$ estimates into a single value per PDF set that properly takes
into account the experimental and theoretical uncertainties and their correlations. The final $\alpS(m_{\PZ})$ values derived are presented
and discussed in Section~\ref{sec:5}. The work is summarised in Section~\ref{sec:6}.

\section{The CMS detector}
\label{sec:exp}

The results presented here are based on a phenomenological study of \PWpm and \PZ boson fiducial cross sections measured by the CMS
experiment in $\Pp\Pp$ collisions at centre-of-mass (c.m.) energies of $\sqrts = 7$ and 8\TeV with integrated luminosities of 38.0 and 18.2\pbinv,
respectively~\cite{CMS:2011aa,Chatrchyan:2014mua}. The experimental and theoretical EW boson production cross sections quoted in the whole paper
are to be understood as multiplied by their associated leptonic branching fractions, but for simplicity are referred to as ``cross sections'' hereafter.
The final states of interest are those with decay charged leptons (electrons or muons) passing the acceptance criteria listed in Table~\ref{tab:data}.

The central feature of the CMS apparatus is a superconducting solenoid, of 6\unit{m} internal diameter, providing a magnetic field of 3.8\unit{T}.
Within the field volume are a silicon pixel and strip tracker, a crystal electromagnetic calorimeter (ECAL), and a brass and scintillator hadron
calorimeter. Electrons with $\pt>25\GeV$ are identified as clusters of energy deposits in the ECAL matched to tracks measured with the silicon tracker.
The ECAL fiducial region is defined by $\abs{\eta}<1.44$ (barrel) or $1.57<\abs{\eta}<2.5$ (endcap), where $\eta$ is the pseudorapidity of the energy cluster.
Muons are measured in gas-ionisation detectors embedded in the steel flux-return yoke of the magnet.
Muons with $\pt>20$ or 25\GeV and $\abs{\eta}<2.1$ are selected in the analyses.
Details of the CMS detector and its performance can be found elsewhere~\cite{Chatrchyan:2008aa}.

\begin{table}[htpb!]
\topcaption{Summary of the twelve \PWpm and \PZ boson production cross sections,
along with their individual (and total, added in quadrature) uncertainties, measured with the indicated fiducial selection criteria
on the transverse momentum ($\pt^\ell$) and pseudorapidity ($\eta^\ell$), in the electron ($\PWpm_{\Pe}$, $\PZ_{\Pe}$) and muon
($\PWpm_{\PGm}$, $\PZ_{\PGm}$) final states, in $\Pp\Pp$ collisions at $\sqrts = 7$ and 8\TeV~\cite{CMS:2011aa,Chatrchyan:2014mua}.}
\centering
\cmsTable{
\begin{tabular}{ll}\hline
Measurement & Fiducial cross section  \\\hline
$\Pp\Pp$ at $\sqrts = 7\TeV$~\cite{CMS:2011aa} & \\
$\PW^{+}_{\Pe}$, $\pt^{\Pe} > 25\GeV$, $\abs{\eta^{\Pe}} < 2.5$ & $3404 \pm 12\stat \pm 67{\syst} \pm 136\lum \unit{pb} = 3404 \pm 152\unit{pb}$ \\
$\PW^{-}_{\Pe}$, $\pt^{\Pe} > 25\GeV$, $\abs{\eta^{\Pe}} < 2.5$ & $2284 \pm 10\stat \pm 43{\syst} \pm 91\lum \unit{pb} = 2284 \pm 101\unit{pb}$ \\
$\PZ_{\Pe}$, $\pt^{\Pe} > 25\GeV$, $\abs{\eta^{\Pe}} < 2.5$, $ 60 < \mZ < 120\GeV$ & $452 \pm 5\stat \pm 10{\syst} \pm 18\lum \unit{pb} = 452 \pm 21\unit{pb}$\\
$\PW^{+}_{\PGm}$, $\pt^{\PGm} > 25\GeV$, $\abs{\eta^{\PGm}} < 2.1$  & $2815 \pm 9\stat \pm 42{\syst} \pm 113\lum \unit{pb} = 2815 \pm 121\unit{pb}$ \\
$\PW^{-}_{\PGm}$, $\pt^{\PGm} > 25\GeV$, $\abs{\eta^{\PGm}} < 2.1$  & $1921 \pm 8\stat \pm 27{\syst} \pm 77\lum \unit{pb} = 1921 \pm 82\unit{pb}$ \\
$\PZ_{\PGm}$, $\pt^{\PGm} > 20\GeV$, $\abs{\eta^{\PGm}} < 2.1$, $ 60 < \mZ < 120\GeV$ & $396 \pm 3\stat \pm 7{\syst} \pm 16\lum \unit{pb} = 396 \pm 18\unit{pb}$  \\[\cmsTabSkip]
$\Pp\Pp$ at $\sqrts = 8\TeV$~\cite{Chatrchyan:2014mua} & \\
$\PW^{+}_{\Pe}$, $\pt^{\Pe} > 25\GeV$, $\abs{\eta^{\Pe}} < 1.44$, $1.57 < \abs{\eta^{\Pe}} < 2.5$ & $3540 \pm 20\stat \pm 110{\syst} \pm 90\lum \unit{pb} = 3540 \pm 140\unit{pb}$ \\
$\PW^{-}_{\Pe}$, $\pt^{\Pe} > 25\GeV$, $\abs{\eta^{\Pe}} < 1.44$, $1.57 < \abs{\eta^{\Pe}} < 2.5$  & $2390 \pm 10\stat \pm 60{\syst} \pm 60\lum \unit{pb} = 2390 \pm 90\unit{pb}$ \\
$\PZ_{\Pe}$, $\pt^{\Pe} > 25\GeV$, $\abs{\eta^{\Pe}} < 1.44$, $1.57 < \abs{\eta^{\Pe}} < 2.5$, $ 60 < \mZ < 120\GeV$ & $450 \pm 10\stat \pm 10{\syst} \pm 10\lum \unit{pb} = 450 \pm 20\unit{pb}$  \\
$\PW^{+}_{\PGm}$, $\pt^{\PGm} > 25\GeV$, $\abs{\eta^{\PGm}} < 2.1$ & $3100 \pm 10\stat \pm 40{\syst} \pm 80\lum \unit{pb} = 3100 \pm 90\unit{pb}$ \\
$\PW^{-}_{\PGm}$, $\pt^{\PGm} > 25\GeV$, $\abs{\eta^{\PGm}} < 2.1$ & $2240 \pm 10\stat \pm 20{\syst} \pm 60\lum \unit{pb} = 2240 \pm 60\unit{pb}$ \\
$\PZ_{\PGm}$, $\pt^{\PGm} > 25\GeV$, $\abs{\eta^{\PGm}} < 2.1$, $ 60 < \mZ < 120\GeV$  & $400 \pm 10\stat \pm 10{\syst} \pm 10\lum \unit{pb} = 400 \pm 20\unit{pb}$  \\ \hline
\end{tabular}
}
\label{tab:data}
\end{table}

\section{Theoretical calculations}
\label{sec:thsetup}

According to the pQCD factorisation theorem~\cite{Collins:1989gx}, the cross section for the production of a heavy elementary particle
in $\Pp\Pp$ collisions can be calculated through the convolution of matrix elements for the relevant parton-parton subprocesses,
computed at a given order in the $\alpS$ expansion evaluated at a renormalisation scale $\PGm_\textsc{r}$, and a universal
nonperturbative part describing the parton density at the factorisation energy scale $\PGm_\textsc{f}$ and parton fractional
momentum $x_i$ in the proton. The production cross section of an EW boson can be written
\begin{linenomath*}
\begin{equation*}
\sigma(\Pp\Pp\to \PV+\text{X}) = \int\int \rd x_1 \rd x_2\,f_1(x_1,\PGm_\textsc{f})f_2(x_2,\PGm_\textsc{f})\,\left[\hat{\sigma}_{\textsc{lo}}+
\hat{\sigma}_{\textsc{nlo}}(\alpS(\PGm_\textsc{r}))+\hat{\sigma}_{\textsc{nnlo}}(\alpS(\PGm_\textsc{r})) + \cdots \right],
\end{equation*}
\end{linenomath*}
where the functions $f_i$ represent the PDFs of each proton, determined from experimental data, and the expression in brackets
is the perturbative expansion of the underlying partonic cross sections $\hat{\sigma}$.
At hadron colliders, the LO production of \PWpm and \PZ bosons involves the annihilation of a
quark-antiquark pair of the same ($\qqbar\to \PZ+\text{X}$) or different ($\qqbar'\to \PW+\text{X}$) flavour. At
NLO, the Born terms are supplemented with initial-state real gluon emission, virtual gluon exchange, and contributions from gluon-quark and
gluon-antiquark scattering processes. At NNLO, additional gluon radiation and virtual exchanges further contribute 
to the total cross section~\cite{Hamberg:1990np,Anastasiou:2003ds}.
Although at LO the partonic cross sections are independent of $\alpS$, the vertices of the higher-order terms introduce a dependence on
$\alpS$ that enables the determination of the QCD coupling by comparing high-precision theoretical calculations to the experimental
data. The size of such higher-order corrections~\cite{Poldaru:2019}, encoded in the so-called $K$-factor, amounts to
$K = \sigma_{\textsc{nnlo}}/\sigma_\textsc{lo}\approx1.25$--1.37 as derived with \MCFM~v.8.0~\cite{Boughezal:2016wmq}
for the \PWpm and \PZ cross sections measured at 7 and 8\TeV in the CMS fiducial acceptance, and indicates that 
EW boson production is indeed sensitive to $\alpS(m_{\PZ})$ at NNLO accuracy.

In this work, the NNLO cross sections are computed with the \MCFM code interfaced with \LHAPDF v.6.1.6~\cite{Buckley:2014ana}
to access four different PDFs: CT14~\cite{Dulat:2015mca}, HERAPDF2.0~\cite{Zhang:2015tuh}, MMHT14~\cite{Harland-Lang:2014zoa},
and NNPDF3.0~\cite{Ball:2014uwa}. All these PDFs use as the default central set the one with the QCD coupling constant fixed to $\alpS(m_{\PZ}) = 0.118$ in their global fits of the data, but also provide a variety of alternative sets with their corresponding central values derived for different fixed values of $\alpS(m_{\PZ})$. We note, however, that when the QCD coupling constant is left free in their NNLO PDF fits, the following values are preferred by the different PDF sets: $\alpS(m_{\PZ})$ = 0.115 (CT14)~\cite{Dulat:2015mca}, 0.108 (HERAPDF2.0)~\cite{Zhang:2015tuh}, and 0.1172 (MMHT2014)~\cite{Harland-Lang:2014zoa}.
The HERAPDF2.0 set is obtained from fits to HERA deep inelastic scattering (DIS) data only. The CT14, MMHT14, and NNPDF3.0 global fits have been obtained including
DIS, fixed target, and LHC measurements. These latter PDF sets incorporate one or two \PWpm or \PZ differential CMS distributions
at 7\TeV~\cite{CMS:2011aa} in their global fits, but did not use any of the twelve \textit{absolute} inclusive EW boson cross
sections listed in Table~\ref{tab:data}, and therefore the corresponding values of $\alpS$ extracted here are truly independent
of the data contributing to the extraction of PDF sets themselves.
The so-called $G_{\PGm}$ electroweak scheme, where the input parameters are $\mW$, $\mZ$, and $G_\text{F}$, is used in all the predictions.
The leptonic \PW and \PZ branching fractions are obtained in \MCFM from the theoretical leptonic width (computed at LO in electroweak accuracy) normalized to the total \PW and \PZ widths experimentally measured~\cite{PDG}.
All numerical results have been obtained using the latest SM parameters for particle masses, widths, and couplings~\cite{PDG}.
For simplicity, the default value of the charm quark mass in \MCFM and NNPDF3.0, $m_\text{\cPqc}=1.275\GeV$, is used for all the
calculations---rather than the preferred values of the other PDF sets: $m_\text{\cPqc}=1.3\GeV$ (CT14), $1.43\GeV$ (HERAPDF2.0),
$1.4\GeV$ (MMHT14)---because the computed cross sections vary only by a few per mille, within the \MCFM numerical uncertainty.

The default renormalisation and factorisation scales are set to the corresponding EW boson mass for each process,
$\PGm = \PGm_\textsc{f} = \PGm_\textsc{r} = \mW,\mZ$.
For all PDF sets, we computed the NNLO cross sections at various $\alpS(m_{\PZ})$ values over the range [0.115--0.121].
For NNPDF3.0, the available values are $\alpS(m_{\PZ}) = 0.115$, 0.117, 0.118, 0.119, and 0.121; and for the other PDF sets they are
$\alpS(m_{\PZ}) = 0.115$, 0.116, 0.117, 0.118, 0.119, 0.120, and 0.121. Technically, the central sets selected via \LHAPDF for 
this study are: CT14nnlo\_as\_0$iii$ (for $iii=115$--121), HERAPDF20\_NNLO\_ALPHAS\_$iii$ (for $iii=115$--121), MMHT2014nnlo\_asmzlargerange 
(with $\alpS(m_{\PZ})=0.115,...,0.121$ grids), and NNPDF30\_nnlo\_as\_0$iii$ (with $iii=115$--121). 
For the PDF uncertainties, only NNPDF3.0 provides independent replicas for each $\alpS(m_{\PZ})$ set, 
which we use in our calculations and uncertainties propagation, whereas the rest of PDFs use the same eigenvalues corresponding 
to the set determined with $\alpS(m_{\PZ}) = 0.118$. Calculations are carried out implementing the fiducial
selection criteria for the final-state charged leptons corresponding to each of the six different
measurements ($\PW^{+}_{\Pe}$, $\PW^{-}_{\Pe}$, $\PZ_{\Pe}$, $\PW^{+}_{\PGm}$, $\PW^{-}_{\PGm}$, $\PZ_{\PGm}$)
at $\sqrts = 7$ and 8\TeV listed in Table~\ref{tab:data}, thereby providing altogether twelve theoretical
cross sections per PDF that can be used to individually extract $\alpS(m_{\PZ})$.

The PDF uncertainties of the theoretical fiducial cross sections are obtained by taking into account the different eigenvector sets,
or replicas, that come with each of the PDFs. We use the ``official'' prescriptions of each PDF set to compute their associated uncertainties.
More specifically, the PDF uncertainties are calculated from the cross sections obtained with the central PDF member ($\sigma_0$)
and with the rest of eigenvalues or replicas ($\sigma_i$) as follows:
\begin{itemize}
\item For CT14, the uncertainty eigenvectors are considered in pairs from the $i = 1$--56 PDF members.
The largest positive and negative differences from each pair are summed quadratically to obtain the corresponding positive and negative PDF uncertainties:
\begin{equation*}
\Delta \sigma_{\pm} = \sqrt{\sum_{j=1}^{28} \max\left( \pm(\sigma_{2j-1}-\sigma_0),\, \pm(\sigma_{2j}-\sigma_0),\,0 \right)^2}.
\end{equation*}
The CT14 PDF set results in asymmetric uncertainties interpreted as a 90\% confidence level interval. To convert those
to one standard deviation, as for the rest of PDF sets, they are divided by a factor of $\sqrt2\erf^{-1}(0.9)\approx1.645$.
\item For HERAPDF2.0, a first asymmetric uncertainty is derived from the so-called 'EIG' (experimental uncertainties) PDF members,
and a second one from the $i = 1$--10 'VAR' (variation) members, as for CT14.
A third asymmetric uncertainty is taken from the $i = 11$--13 VAR members, as the maximum positive and negative differences
$\Delta \sigma_i$ with respect to $\sigma_0$.
Finally, all positive and negative uncertainties are separately added quadratically to get the final uncertainties.
\item For MMHT14, uncertainties are obtained from its corresponding 50 eigenvalues as done for CT14.
\item For NNPDF3.0, the average cross section $\hat{\sigma}$ from replica members $i = 1$--100 is calculated first, and the associated
standard deviation, $\sqrt{\sum_{i=1}^{100} (\sigma_i-\hat{\sigma})^2/99}$, is taken as the symmetric PDF uncertainty.
\end{itemize}

To determine the scale uncertainty associated with missing corrections beyond the NNLO accuracy, the \MCFM cross sections are
recalculated for each PDF and measurement using factorisation and renormalisation scales varied within factors of two, such that the ratio of the
two scales is not less than 0.5 or more than 2. This gives seven combinations: $(\PGm_\textsc{f},\PGm_\textsc{r})$, $(\PGm_\textsc{f}/2,\PGm_\textsc{r}/2)$,
$(\PGm_\textsc{f}/2,\PGm_\textsc{r})$, $(\PGm_\textsc{f},\PGm_\textsc{r}/2)$, $(2\PGm_\textsc{f},\PGm_\textsc{r})$,
$(\PGm_\textsc{f},2\PGm_\textsc{r})$, $(2\PGm_\textsc{f},2\PGm_\textsc{r})$. The largest and smallest cross sections of
the seven combinations are determined, and the
scale uncertainty is taken as half the difference of the extremal values. The scale variation uncertainties amount to
0.5--1\% of the theoretical cross sections.

Since \MCFM does not include EW corrections, arising from additional \PWpm, \PZ, and/or photons exchanged and/or
radiated in the partonic process, those are computed separately. For this purpose, the \MCSANC v.1.01 code~\cite{Bondarenko:2013nu} is used.
For $\Pe^{\pm}$ final states, we follow the ``calorimetric'' prescription, proposed in theoretical
\PWpm and \PZ boson production cross section benchmarking studies at the LHC~\cite{Alioli:2016fum},
and recombine any radiated photon with the $\Pe^{\pm}$ if their relative distance in the pseudorapidity-azimuth plane is
$\Delta R = \sqrt{(\eta^{\Pe}-\eta^\gamma)^2+(\phi^{\Pe}-\phi^\gamma)^2}< 0.1$.
For $\PGm^{\pm}$ final states, we use directly the ``bare'' \MCSANC cross section. We run \MCSANC at NLO with EW corrections on and off,
and compute the corresponding multiplicative factor $K_\textsc{EW}=\sigma(\textsc{nlo,ew}\,\text{on})/\sigma(\textsc{nlo,ew}\,\text{off})$,
which is used to correct the pQCD \MCFM results. The EW corrections, in the range of 1--4\%, are all negative, \ie\ they reduce the overall
cross section with respect to the pure pQCD result. Since the EW corrections are small, their associated uncertainties
are neglected hereafter because they would propagate into the final computed \PWpm and \PZ\ cross section at a few per mille level,
below the numerical uncertainty of the \MCFM calculation. Subtracting the EW corrections, rather than applying them multiplicatively via a
$K_\textsc{EW}$ factor as done here, gives consistent results within the (neglected) per mille uncertainties.
For simplicity, in the \MCSANC calculations, the electron pseudorapidity range $1.44 < \abs{\eta} < 1.57$ (excluded in the actual measurements)
is also included, since we are interested in the relative correction, this small range (present in both the numerator and denominator of the correcting factor)
does not affect the $K_\textsc{EW}$ ratio.
The roles of photon-induced contributions and of mixed QCD$\oplus$QED NLO corrections to Drell--Yan processes in $\Pp\Pp$ collisions have
been computed in Refs.~\cite{Bertone:2017bme} and~\cite{deFlorian:2018wcj}, respectively. The impact of such corrections to the inclusive
\PWpm and \PZ\ cross sections is at a few per mille level, and also neglected here.

All the relevant sources of uncertainties in the \PWpm and \PZ boson cross sections are summarised in Table~\ref{tab:uncertainties_overview}.
The largest experimental and theoretical uncertainties come from the integrated luminosity and PDF knowledge, respectively.
Each calculated cross section has a numerical accuracy, as reported by \MCFM, in the range of 0.2--0.6\%. Such an uncertainty is
commensurate with the typical differences encountered when computing NNLO \PWpm and \PZ boson cross sections with different pQCD codes
that implement higher-order virtual-real corrections with various methods~\cite{Alioli:2016fum}.

\begin{table}
\topcaption{Summary of the typical experimental and theoretical uncertainties in the \PWpm and \PZ boson production cross sections,
and their degree of correlation (details are provided in Section~\ref{sec:combiningEstimates}).
\label{tab:uncertainties_overview}}
\centering
\begin{tabular}{lll}\\\hline
Uncertainties & & Degree of correlation \\\hline
Experimental: & & \\
Integrated luminosity & 2--4\% & fully correlated at a given c.m.\ energy \\
Systematic & 1--3\% & partially correlated \\
Statistical & 0.5--2\% & uncorrelated \\[\cmsTabSkip]
Theoretical: & &  \\
PDF & 1--4\% & partially correlated within each PDF set \\
Theoretical scale & 0.3--1.3\% & partially correlated \\
\MCFM statistical numerical & 0.2--0.6\% & uncorrelated \\\hline
\end{tabular}
\end{table}

\section{EW boson fiducial cross sections: data versus theory}
\label{sec:cross_sections}

All the experimental and theoretical fiducial cross sections
for \PWp, \PWm, and \PZ production in $\Pp\Pp$ collisions are given in Tables~\ref{tab:CMS7_x} and \ref{tab:CMS8_x} for
7 and 8\TeV, respectively. For each measurement, the fiducial cross section definition and the experimental
result are listed along with their uncertainties from the different sources listed in Table~\ref{tab:uncertainties_overview}.
The theoretical \MCFM predictions computed with all four PDF sets for their preferred default $\alpS(m_{\PZ})=0.118$ value are listed
including their associated PDF, $\alpS$ (obtained, as described in Section~\ref{sec:uncert_propag}, from the cross section 
change when $\alpS(m_{\PZ})$ is modified by $\pm 0.001$), and scale uncertainties. 
For each system, the NLO \MCSANC EW correction factors (absolute and relative) are also listed.
For the results at $\sqrts = 8\TeV$, the theoretical result obtained with the alternative \FEWZ NNLO pQCD calculator~\cite{Li:2012wna},
using the MSTW2008 PDF set~\cite{Martin:2009iq} as provided in the original Ref.~\cite{Chatrchyan:2014mua},
is also listed to show the very similar theoretical predictions expected with an alternative NNLO code and a pre-LHC PDF set.

\begin{table}[htbp!]
\topcaption{Experimental and theoretical fiducial cross sections for \PWpm and \PZ production
in $\Pp\Pp$ collisions at $\sqrts = 7\TeV$, with the uncertainty sources listed in Table~\ref{tab:uncertainties_overview}.
The NNLO pQCD results are obtained with \MCFM for $\alpS(m_{\PZ}) = 0.118$ using the CT14, HERAPDF2.0, MMHT14, and NNPDF3.0 PDF sets.
(The quoted $\alpS$ uncertainties are derived from the cross section changes when $\alpS(m_{\cPZ})$ is modified by $\pm 0.001$).
The NLO EW corrections are computed with \MCSANC~\label{tab:CMS7_x}}.
\centering
\cmsTable{
\renewcommand*{\arraystretch}{1.2}
\begin{tabular}{ll}\hline
System & Fiducial cross section  \\\hline
$\Pp\Pp\to \PWp(\Pe^+\Pgn)+\text{X}$, $\sqrts$ = 7\TeV ($\pt^{\Pe} > 25\GeV$, $\abs{\eta^{\Pe}} < 2.5$) & \\
Measurement~\cite{CMS:2011aa}           & $3404 \pm 12\stat \pm 67{\syst} \pm 136\lum \unit{pb}$ \\
\MCFM (NNLO, CT14)                     & $3361\,^{+93}_{-94}{\,\text{(PDF)}} \pm 30{\,(\alpS)} \pm 49\,\text{(scale)} \pm 18\stat \unit{pb}$ \\
\MCFM (NNLO, HERAPDF2.0)               & $3574\,^{+63}_{-94}{\,\text{(PDF)}} \pm 19{\,(\alpS)} \pm 33\,\text{(scale)} \pm 23\stat \unit{pb}$ \\
\MCFM (NNLO, MMHT14)                   & $3407\,^{+92}_{-74}{\,\text{(PDF)}} \pm 37{\,(\alpS)} \pm 31\,\text{(scale)} \pm 18\stat \unit{pb}$ \\
\MCFM (NNLO, NNPDF3.0)                 & $3345 \pm 70{\,\text{(PDF)}} \pm 32{\,(\alpS)} \pm 29\,\text{(scale)} \pm 18\stat \unit{pb}$ \\
\MCSANC (NLO EW correction, NNPDF3.0) & $-36 \unit{pb} \; (-1.1\%)$ \\[\cmsTabSkip]
$\Pp\Pp\to \PWm(\Pe^-\Pagn)+\text{X}$, $\sqrts$ = 7\TeV ($\pt^{\Pe} > 25\GeV$, $\abs{\eta^{\Pe}} < 2.5$) & \\
Measurement~\cite{CMS:2011aa}           & $2284 \pm 10\stat \pm 43{\syst} \pm 91\lum \unit{pb}$ \\
\MCFM (NNLO, CT14)                     & $2235\,^{+66}_{-57}{\,\text{(PDF)}} \pm 19{\,(\alpS)} \pm 19\,\text{(scale)} \pm 7\stat \unit{pb}$ \\
\MCFM (NNLO, HERAPDF2.0)               & $2319\,^{+21}_{-51}{\,\text{(PDF)}} \pm 8{\,(\alpS)} \pm 19\,\text{(scale)} \pm 7\stat \unit{pb}$ \\
\MCFM (NNLO, MMHT14)                   & $2248\,^{+28}_{-62}{\,\text{(PDF)}} \pm 23{\,(\alpS)} \pm 17\,\text{(scale)} \pm 7\stat \unit{pb}$ \\
\MCFM (NNLO, NNPDF3.0)                 & $2192 \pm 47{\,\text{(PDF)}} \pm 16{\,(\alpS)} \pm 16\,\text{(scale)} \pm 7\stat \unit{pb}$ \\
\MCSANC (NLO EW correction, NNPDF3.0) & $-24 \unit{pb} \; (-1.1\%)$ \\[\cmsTabSkip]
$\Pp\Pp\to  \PZ(\Pe^+\Pe^-)+\text{X}$, $\sqrts$ = 7\TeV ($\pt^{\Pe} > 25\GeV$, $\abs{\eta^{\Pe}} < 2.5$, $60 < \mZ < 120\GeV$) & \\
Measurement~\cite{CMS:2011aa}           & $452 \pm 5\stat \pm 10{\syst} \pm 18\lum \unit{pb}$ \\
\MCFM (NNLO, CT14)                     & $430\,^{+11}_{-13}{\,\text{(PDF)}} \pm 4{\,(\alpS)} \pm 2\,\text{(scale)} \pm 1\stat \unit{pb}$ \\
\MCFM (NNLO, HERAPDF2.0)               & $444\,^{+4}_{-12}{\,\text{(PDF)}} \pm 2{\,(\alpS)} \pm 2\,\text{(scale)} \pm 1\stat \unit{pb}$ \\
\MCFM (NNLO, MMHT14)                   & $433\,^{+6}_{-10}{\,\text{(PDF)}} \pm 5{\,(\alpS)} \pm 3\,\text{(scale)} \pm 1\stat \unit{pb}$ \\
\MCFM (NNLO, NNPDF3.0)                 & $421 \pm 9{\,\text{(PDF)}} \pm 3{\,(\alpS)} \pm 2\,\text{(scale)} \pm 1\stat \unit{pb}$ \\
\MCSANC (NLO EW correction, NNPDF3.0) & $-12 \unit{pb} \; (-2.6\%)$ \\[\cmsTabSkip]
$\Pp\Pp\to \PWp(\PGm^+\nu)+\text{X}$, $\sqrts$ = 7\TeV ($\pt^{\PGm} > 25\GeV$, $\abs{\eta^{\PGm}} < 2.1$) & \\
Measurement~\cite{CMS:2011aa}           & $2815 \pm 9\stat \pm 42{\syst} \pm 113\lum \unit{pb}$ \\
\MCFM (NNLO, CT14)                     & $2827\,^{+65}_{-110}{\,\text{(PDF)}} \pm 29{\,(\alpS)} \pm 21\,\text{(scale)} \pm 13\stat \unit{pb}$ \\
\MCFM (NNLO, HERAPDF2.0)               & $2976\,^{+42}_{-118}{\,\text{(PDF)}} \pm 16{\,(\alpS)} \pm 37\,\text{(scale)} \pm 15\stat \unit{pb}$ \\
\MCFM (NNLO, MMHT14)                   & $2833\,^{+63}_{-90}{\,\text{(PDF)}} \pm 29{\,(\alpS)} \pm 17\,\text{(scale)} \pm 16\stat \unit{pb}$ \\
\MCFM (NNLO, NNPDF3.0)                 & $2806 \pm 62{\,\text{(PDF)}} \pm 26{\,(\alpS)} \pm 29\,\text{(scale)} \pm 15\stat \unit{pb}$ \\
\MCSANC (NLO EW correction, NNPDF3.0) & $-64 \unit{pb} \; (-2.2\%)$ \\[\cmsTabSkip]
$\Pp\Pp\to \PWm(\PGm^-\Pagn)+\text{X}$, $\sqrts$ = 7\TeV ($\pt^{\PGm} > 25\GeV$, $\abs{\eta^{\PGm}} < 2.1$) & \\
Measurement~\cite{CMS:2011aa}           & $1921 \pm 8\stat \pm 27{\syst} \pm 77\lum \unit{pb}$ \\
\MCFM (NNLO, CT14)                     & $1915\,^{+43}_{-68}{\,\text{(PDF)}} \pm 19{\,(\alpS)} \pm 16\,\text{(scale)} \pm 6\stat \unit{pb}$ \\
\MCFM (NNLO, HERAPDF2.0)               & $1976\,^{+33}_{-29}{\,\text{(PDF)}} \pm 8{\,(\alpS)} \pm 19\,\text{(scale)} \pm 6\stat \unit{pb}$ \\
\MCFM (NNLO, MMHT14)                   & $1937\,^{+33}_{-41}{\,\text{(PDF)}} \pm 20{\,(\alpS)} \pm 17\,\text{(scale)} \pm 6\stat \unit{pb}$ \\
\MCFM (NNLO, NNPDF3.0)                 & $1877 \pm 40{\,\text{(PDF)}} \pm 13{\,(\alpS)} \pm 17\,\text{(scale)} \pm 6\stat \unit{pb}$ \\
\MCSANC (NLO EW correction, NNPDF3.0) & $-42 \unit{pb} \; (-2.2\%)$ \\[\cmsTabSkip]
$\Pp\Pp\to \PZ(\PGm^+\PGm^-)+\text{X}$, $\sqrts$ = 7\TeV ($\pt^{\PGm} > 20\GeV$, $\abs{\eta^{\PGm}} < 2.1$, $60 < \mZ < 120\GeV$) & \\
Measurement~\cite{CMS:2011aa}           & $396 \pm 3\stat \pm 7{\syst} \pm 16\lum \unit{pb}$ \\
\MCFM (NNLO, CT14)                     & $380\,^{+7}_{-16}{\,\text{(PDF)}} \pm 3{\,(\alpS)} \pm 2\,\text{(scale)} \pm 1\stat \unit{pb}$ \\
\MCFM (NNLO, HERAPDF2.0)               & $392\,^{+6}_{-6}{\,\text{(PDF)}} \pm 2{\,(\alpS)} \pm 2\,\text{(scale)} \pm 1\stat \unit{pb}$ \\
\MCFM (NNLO, MMHT14)                   & $382\,^{+11}_{-4}{\,\text{(PDF)}} \pm 4{\,(\alpS)} \pm 2\,\text{(scale)} \pm 1\stat \unit{pb}$ \\
\MCFM (NNLO, NNPDF3.0)                 & $373 \pm 8{\,\text{(PDF)}} \pm 3{\,(\alpS)} \pm 2\,\text{(scale)} \pm 1\stat \unit{pb}$ \\
\MCSANC (NLO EW correction, NNPDF3.0) & $-14 \unit{pb} \; (-3.9\%)$ \\ \hline
\end{tabular}
}
\end{table}

\begin{table}[htbp!]
\topcaption{Experimental and theoretical fiducial cross sections for \PWpm and \PZ production
in $\Pp\Pp$ collisions at $\sqrts = 8\TeV$,
with the uncertainty sources listed in Table~\ref{tab:uncertainties_overview}.
The NNLO pQCD results are obtained with \MCFM for $\alpS(m_{\PZ}) = 0.118$ using the CT14, HERAPDF2.0, MMHT14,
and NNPDF3.0 PDF sets, as well as with \FEWZ using the MSTW2008 PDF.
(The quoted $\alpS$ uncertainties are derived from the cross section changes when $\alpS(m_{\cPZ})$ is modified by $\pm 0.001$).
The NLO EW corrections are computed with \MCSANC~\label{tab:CMS8_x}}.
\centering
\cmsTable{
\renewcommand*{\arraystretch}{1.2}
\begin{tabular}{ll}\hline
System & Fiducial cross section  \\\hline
$\Pp\Pp\to \PWp(\Pe^+\nu)+\text{X}$, $\sqrts$ = 8\TeV ($\pt^{\Pe} > 25\GeV$, $\abs{\eta^{\Pe}} < 1.44$, $1.57 < \abs{\eta^{\Pe}} < 2.5$) & \\
Measurement~\cite{Chatrchyan:2014mua}   & $3540 \pm 20\stat \pm 110{\syst} \pm 90\lum \unit{pb}$ \\
\FEWZ (NNLO, MSTW2008)~\cite{Chatrchyan:2014mua}  & $3450 \pm 120 \unit{pb}$ \\
\MCFM (NNLO, CT14)                     & $3522\,^{+113}_{-123}{\,\text{(PDF)}} \pm 23{\,(\alpS)} \pm 35\,\text{(scale)} \pm 21\stat \unit{pb}$ \\
\MCFM (NNLO, HERAPDF2.0)               & $3721\,^{+127}_{-97}{\,\text{(PDF)}} \pm 13{\,(\alpS)} \pm 48\,\text{(scale)} \pm 22\stat \unit{pb}$ \\
\MCFM (NNLO, MMHT14)                   & $3581\,^{+61}_{-137}{\,\text{(PDF)}} \pm 36{\,(\alpS)} \pm 38\,\text{(scale)} \pm 20\stat \unit{pb}$ \\
\MCFM (NNLO, NNPDF3.0)                 & $3515 \pm 75{\,\text{(PDF)}} \pm 34{\,(\alpS)} \pm 42\,\text{(scale)} \pm 20\stat \unit{pb}$ \\
\MCSANC (NLO EW correction, NNPDF3.0) & $-45 \unit{pb} \; (-1.2\%)$ \\[\cmsTabSkip]
$\Pp\Pp\to \PWm(\Pe^-\Pagn)+\text{X}$, $\sqrts$ = 8\TeV ($\pt^{\Pe} > 25\GeV$, $\abs{\eta^{\Pe}} < 1.44$, $1.57 < \abs{\eta^{\Pe}} < 2.5$) & \\
Measurement~\cite{Chatrchyan:2014mua}   & $2390 \pm 10\stat \pm 60{\syst} \pm 60\lum \unit{pb}$ \\
\FEWZ (NNLO, MSTW2008)~\cite{Chatrchyan:2014mua}  & $2380 \pm 90 \unit{pb}$ \\
\MCFM (NNLO, CT14)                     & $2426\,^{+69}_{-61}{\,\text{(PDF)}} \pm 24{\,(\alpS)} \pm 14\,\text{(scale)} \pm 8\stat \unit{pb}$ \\
\MCFM (NNLO, HERAPDF2.0)               & $2513\,^{+51}_{-44}{\,\text{(PDF)}} \pm 11{\,(\alpS)} \pm 21\,\text{(scale)} \pm 10\stat \unit{pb}$ \\
\MCFM (NNLO, MMHT14)                   & $2445\,^{+40}_{-67}{\,\text{(PDF)}} \pm 28{\,(\alpS)} \pm 26\,\text{(scale)} \pm 8\stat \unit{pb}$ \\
\MCFM (NNLO, NNPDF3.0)                 & $2375 \pm 51{\,\text{(PDF)}} \pm 17{\,(\alpS)} \pm 14\,\text{(scale)} \pm 8\stat \unit{pb}$ \\
\MCSANC (NLO EW correction, NNPDF3.0) & $-30 \unit{pb} \; (-1.2\%)$ \\[\cmsTabSkip]
$\Pp\Pp\to \PZ(\Pe^+\Pe^-)+\text{X}$, $\sqrts$ = 8\TeV ($\pt^{\Pe} > 25\GeV$, $\abs{\eta^{\Pe}} < 1.44$, $1.57 < \abs{\eta^{\Pe}} < 2.5$, & \\
 $60 < \mZ < 120\GeV$) & \\
Measurement~\cite{Chatrchyan:2014mua}   & $450 \pm 10\stat \pm 10{\syst} \pm 10\lum \unit{pb}$ \\
\FEWZ (NNLO, MSTW2008)~\cite{Chatrchyan:2014mua}  & $450 \pm 20 \unit{pb}$ \\
\MCFM (NNLO, CT14)                     & $437\,^{+11}_{-15}{\,\text{(PDF)}} \pm 4{\,(\alpS)} \pm 2\,\text{(scale)} \pm 1\stat \unit{pb}$ \\
\MCFM (NNLO, HERAPDF2.0)               & $451\,^{+7}_{-11}{\,\text{(PDF)}} \pm 2{\,(\alpS)} \pm 2\,\text{(scale)} \pm 1\stat \unit{pb}$ \\
\MCFM (NNLO, MMHT14)                   & $441\,^{+11}_{-6}{\,\text{(PDF)}} \pm 5{\,(\alpS)} \pm 3\,\text{(scale)} \pm 1\stat \unit{pb}$ \\
\MCFM (NNLO, NNPDF3.0)                 & $429 \pm 9{\,\text{(PDF)}} \pm 3{\,(\alpS)} \pm 2\,\text{(scale)} \pm 1\stat \unit{pb}$ \\
\MCSANC (NLO EW correction, NNPDF3.0) & $-13 \unit{pb} \; (-2.7\%)$ \\[\cmsTabSkip]
$\Pp\Pp\to \PWp(\PGm^+\nu)+\text{X}$, $\sqrts$ = 8\TeV ($\pt^{\PGm} > 25\GeV$, $\abs{\eta^{\PGm}} < 2.1$) & \\
Measurement~\cite{Chatrchyan:2014mua}   & $3100 \pm 10\stat \pm 40{\syst} \pm 80\lum \unit{pb}$ \\
\FEWZ (NNLO, MSTW2008)~\cite{Chatrchyan:2014mua}  & $3140 \pm 110 \unit{pb}$ \\
\MCFM (NNLO, CT14)                     & $3108\,^{+94}_{-87}{\,\text{(PDF)}} \pm 25{\,(\alpS)} \pm 34\,\text{(scale)} \pm 19\stat \unit{pb}$ \\
\MCFM (NNLO, HERAPDF2.0)               & $3309\,^{+20}_{-153}{\,\text{(PDF)}} \pm 18{\,(\alpS)} \pm 34\,\text{(scale)} \pm 17\stat \unit{pb}$ \\
\MCFM (NNLO, MMHT14)                   & $3148\,^{+67}_{-95}{\,\text{(PDF)}} \pm 33{\,(\alpS)} \pm 29\,\text{(scale)} \pm 15\stat \unit{pb}$ \\
\MCFM (NNLO, NNPDF3.0)                 & $3095 \pm 69{\,\text{(PDF)}} \pm 30{\,(\alpS)} \pm 21\,\text{(scale)} \pm 18\stat \unit{pb}$ \\
\MCSANC (NLO EW correction, NNPDF3.0) & $-77 \unit{pb} \; (-2.4\%)$ \\[\cmsTabSkip]
$\Pp\Pp\to \PWm(\PGm^-\Pagn)+\text{X}$, $\sqrts$ = 8\TeV ($\pt^{\PGm} > 25\GeV$, $\abs{\eta^{\PGm}} < 2.1$) & \\
Measurement~\cite{Chatrchyan:2014mua}   & $2240 \pm 10\stat \pm 20{\syst} \pm 60\lum \unit{pb}$ \\
\FEWZ (NNLO, MSTW2008)~\cite{Chatrchyan:2014mua}  & $2220 \pm 80 \unit{pb}$ \\
\MCFM (NNLO, CT14)                     & $2187\,^{+74}_{-56}{\,\text{(PDF)}} \pm 19{\,(\alpS)} \pm 14\,\text{(scale)} \pm 6\stat \unit{pb}$ \\
\MCFM (NNLO, HERAPDF2.0)               & $2274\,^{+40}_{-20}{\,\text{(PDF)}} \pm 10{\,(\alpS)} \pm 17\,\text{(scale)} \pm 7\stat \unit{pb}$ \\
\MCFM (NNLO, MMHT14)                   & $2200\,^{+42}_{-36}{\,\text{(PDF)}} \pm 23{\,(\alpS)} \pm 20\,\text{(scale)} \pm 7\stat \unit{pb}$ \\
\MCFM (NNLO, NNPDF3.0)                 & $2148 \pm 48{\,\text{(PDF)}} \pm 16{\,(\alpS)} \pm 17\,\text{(scale)} \pm 7\stat \unit{pb}$ \\
\MCSANC (NLO EW correction, NNPDF3.0) & $-47 \unit{pb} \; (-2.1\%)$ \\[\cmsTabSkip]
$\Pp\Pp\to \PZ(\PGm^+\PGm^-)+\text{X}$, $\sqrts$ = 8\TeV ($\pt^{\PGm} > 25\GeV$, $\abs{\eta^{\PGm}} < 2.1$, $ 60 < \mZ < 120\GeV$) & \\
Measurement~\cite{Chatrchyan:2014mua}   & $400 \pm 10\stat \pm 10{\syst} \pm 10\lum \unit{pb}$ \\
\FEWZ (NNLO, MSTW2008)~\cite{Chatrchyan:2014mua}  & $400 \pm 10 \unit{pb}$ \\
\MCFM (NNLO, CT14)                     & $389\,^{+12}_{-12}{\,\text{(PDF)}} \pm 3{\,(\alpS)} \pm 2\,\text{(scale)} \pm 1\stat \unit{pb}$ \\
\MCFM (NNLO, HERAPDF2.0)               & $401\,^{+6}_{-8}{\,\text{(PDF)}} \pm 2{\,(\alpS)} \pm 2\,\text{(scale)} \pm 1\stat \unit{pb}$ \\
\MCFM (NNLO, MMHT14)                   & $391\,^{+11}_{-3}{\,\text{(PDF)}} \pm 4{\,(\alpS)} \pm 2\,\text{(scale)} \pm 1\stat \unit{pb}$ \\
\MCFM (NNLO, NNPDF3.0)                 & $381 \pm 8{\,\text{(PDF)}} \pm 3{\,(\alpS)} \pm 2\,\text{(scale)} \pm 1\stat \unit{pb}$ \\
\MCSANC (NLO EW correction, NNPDF3.0) & $-16 \unit{pb} \; (-3.9\%)$ \\\hline
\end{tabular}
}
\end{table}

\begin{figure}[htpb!]
\centering
\includegraphics[width=0.49\textwidth]{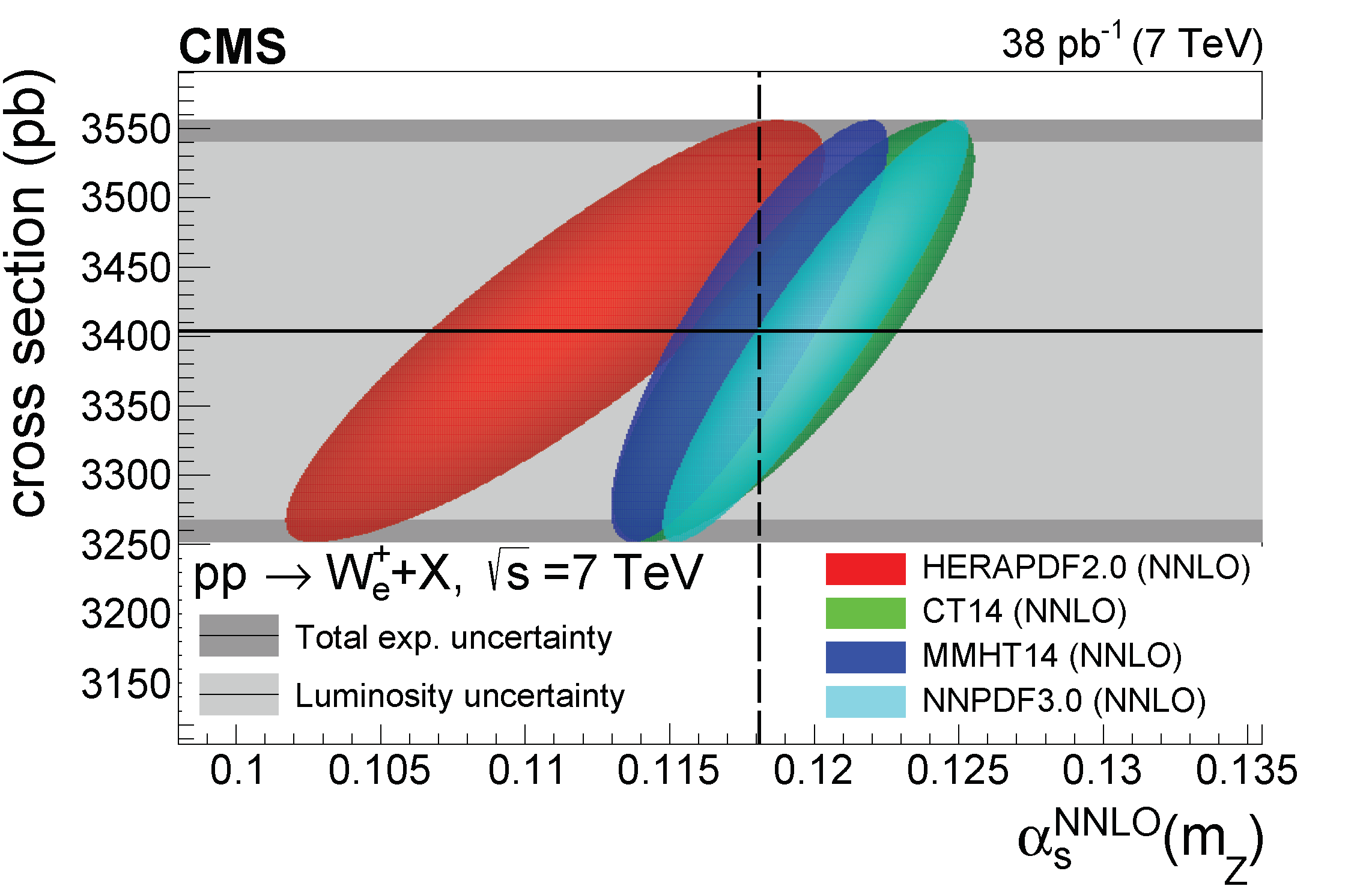}
\includegraphics[width=0.49\textwidth]{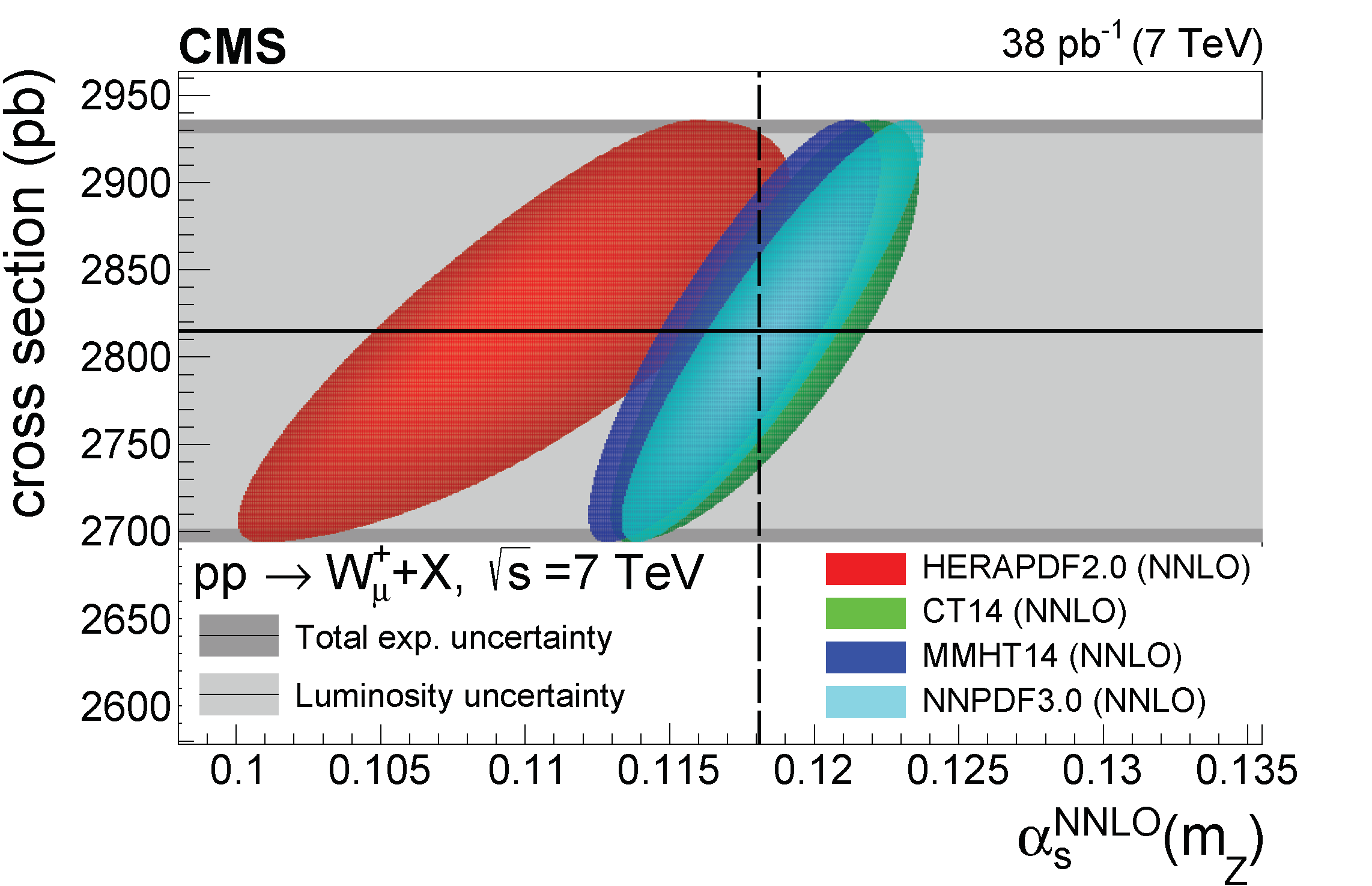}
\includegraphics[width=0.49\textwidth]{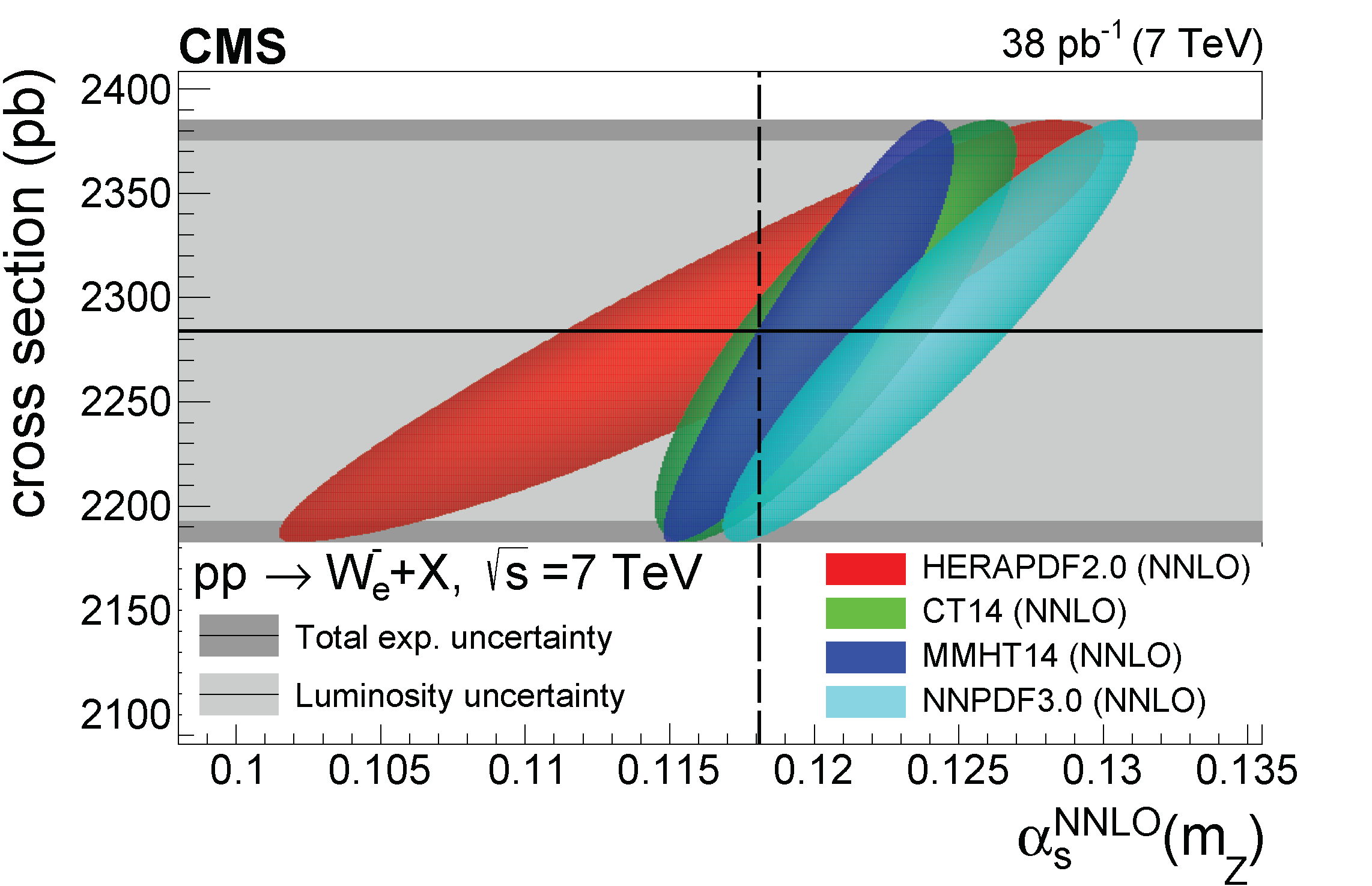}
\includegraphics[width=0.49\textwidth]{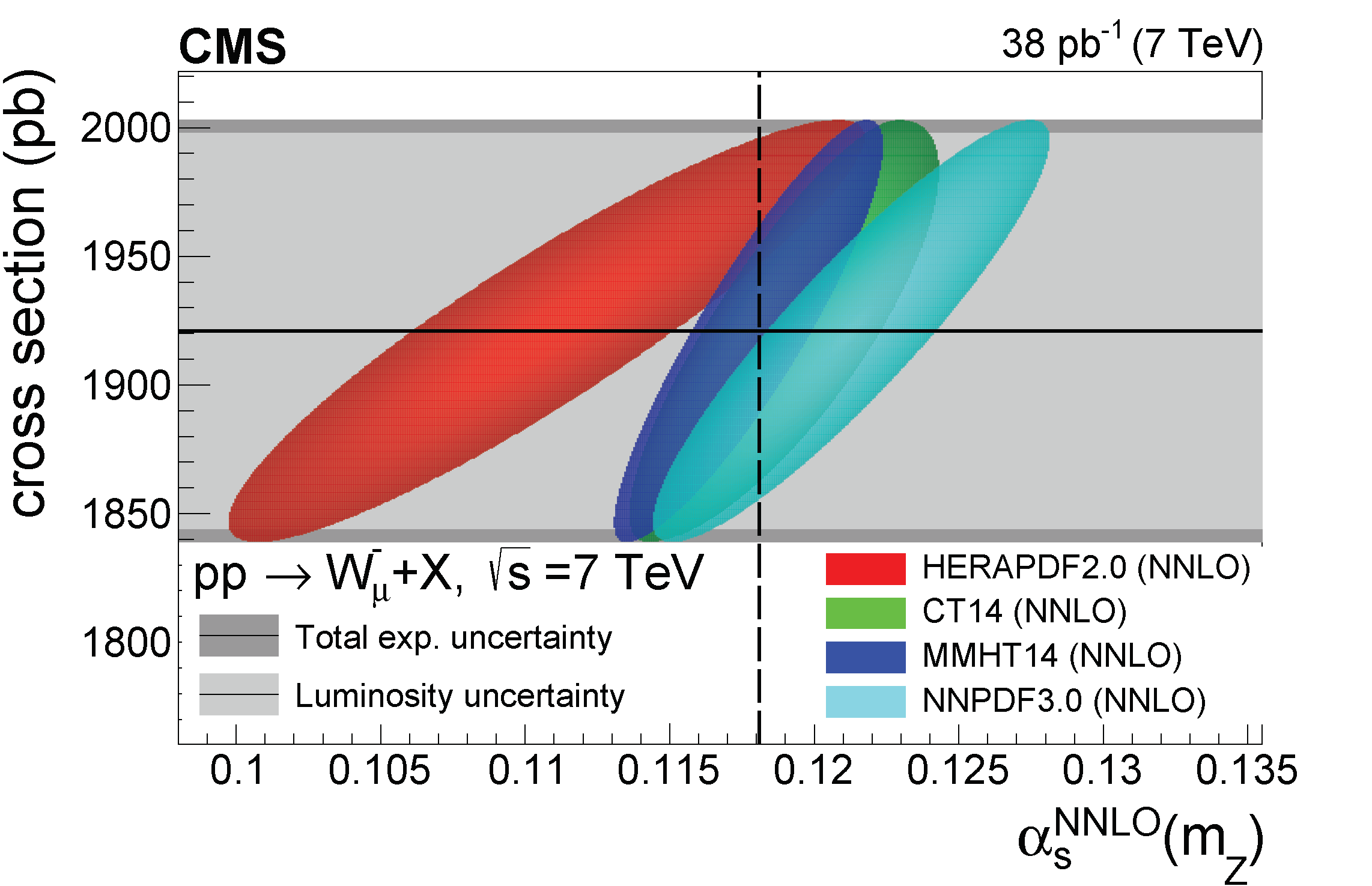}
\includegraphics[width=0.49\textwidth]{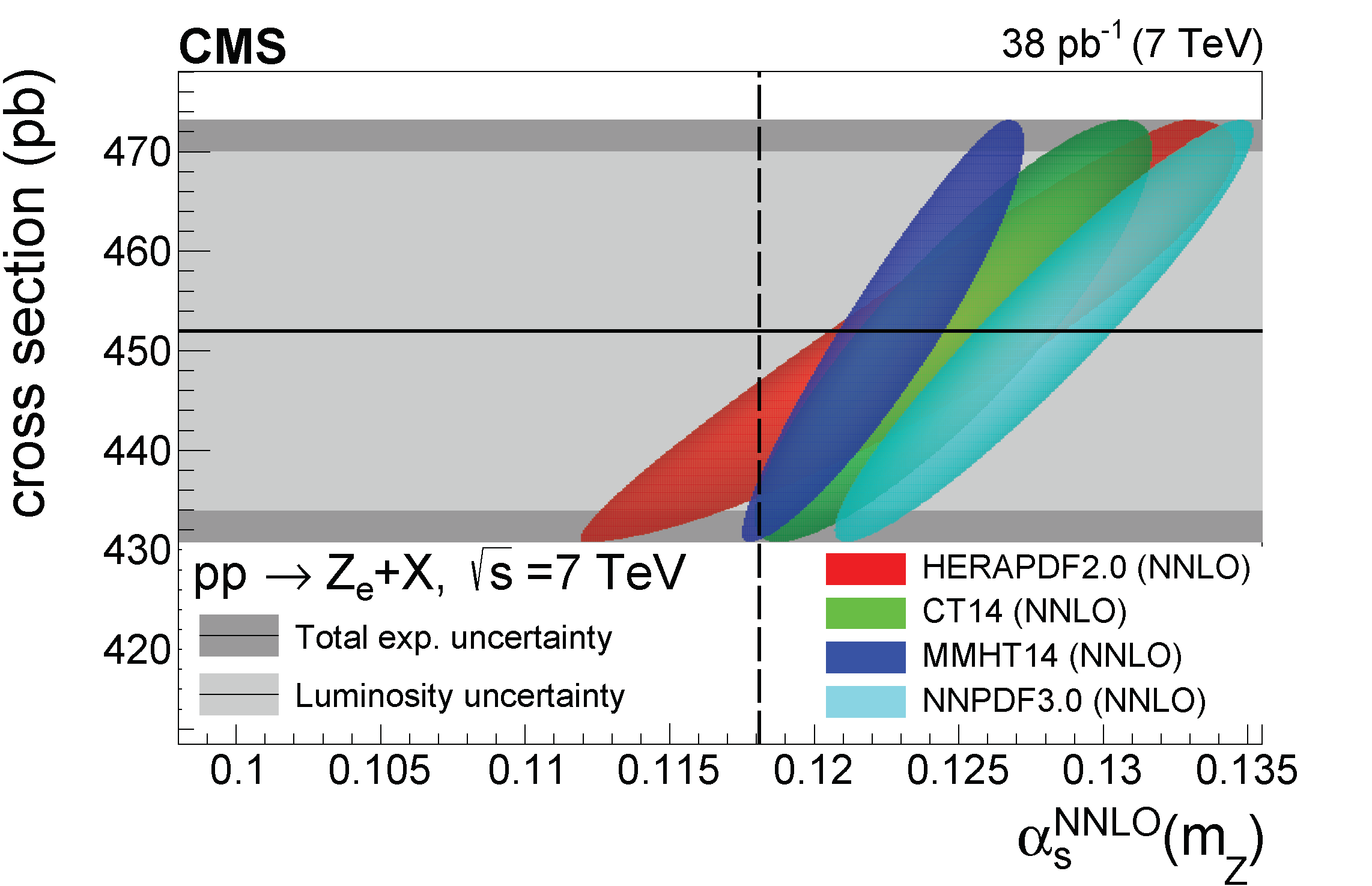}
\includegraphics[width=0.49\textwidth]{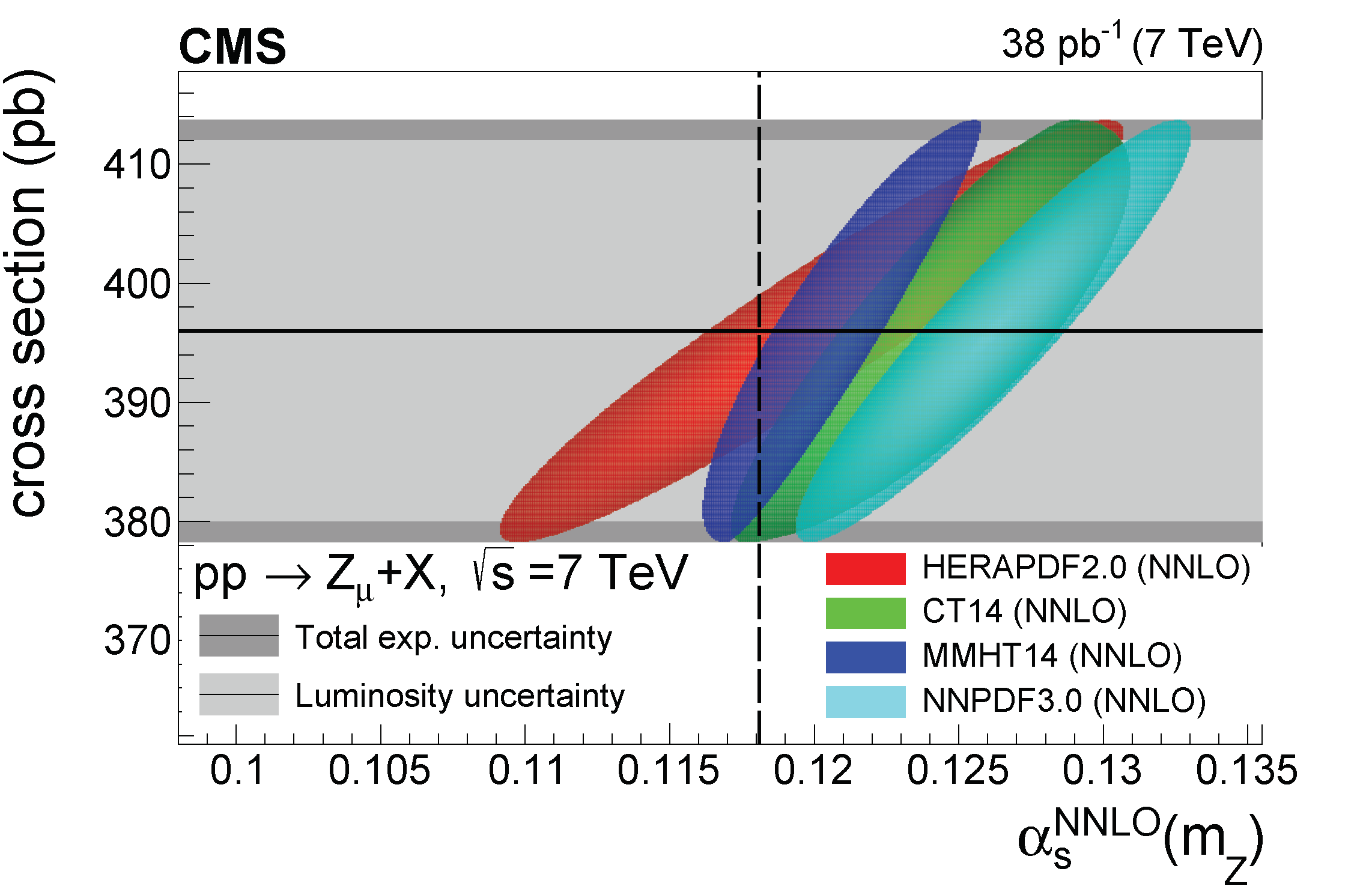}
\caption{Experimental fiducial cross section for the production of $\PW^{+}_{\Pe}$ (upper left) and $\PW^{+}_{\PGm}$ (upper right),
$\PW^{-}_{\Pe}$ (middle left) and $\PW^{-}_{\PGm}$ (middle right), and $\PZ_{\Pe}$ (lower left) and $\PZ_{\PGm}$ (lower right)
in $\Pp\Pp$ collisions at $\sqrts =  7\TeV$ compared to the corresponding joint probability density functions
(elliptical contours, see text) obtained with four different PDFs as a function of $\alpS(m_{\PZ})$ and $\sigma$.
The experimental measurements are plotted as a horizontal black line with the inner grey band indicating the integrated luminosity uncertainty,
and the outer darker band showing all experimental uncertainties added in quadrature. The filled ellipses are obtained from the product of
the probability distributions of
the experimental and theoretical results for each PDF, and represent the two-dimensional one standard deviation.
The points where the filled ellipses cross the vertical dashed line at $\alpS(m_{\PZ}) = 0.118$ indicate the most likely cross section interval
that would be obtained using the baseline QCD coupling constant value of all PDF sets.
\label{fig:fitgraphfirst}}
\end{figure}

\begin{figure}[htpb!]
\centering
\includegraphics[width=0.49\textwidth]{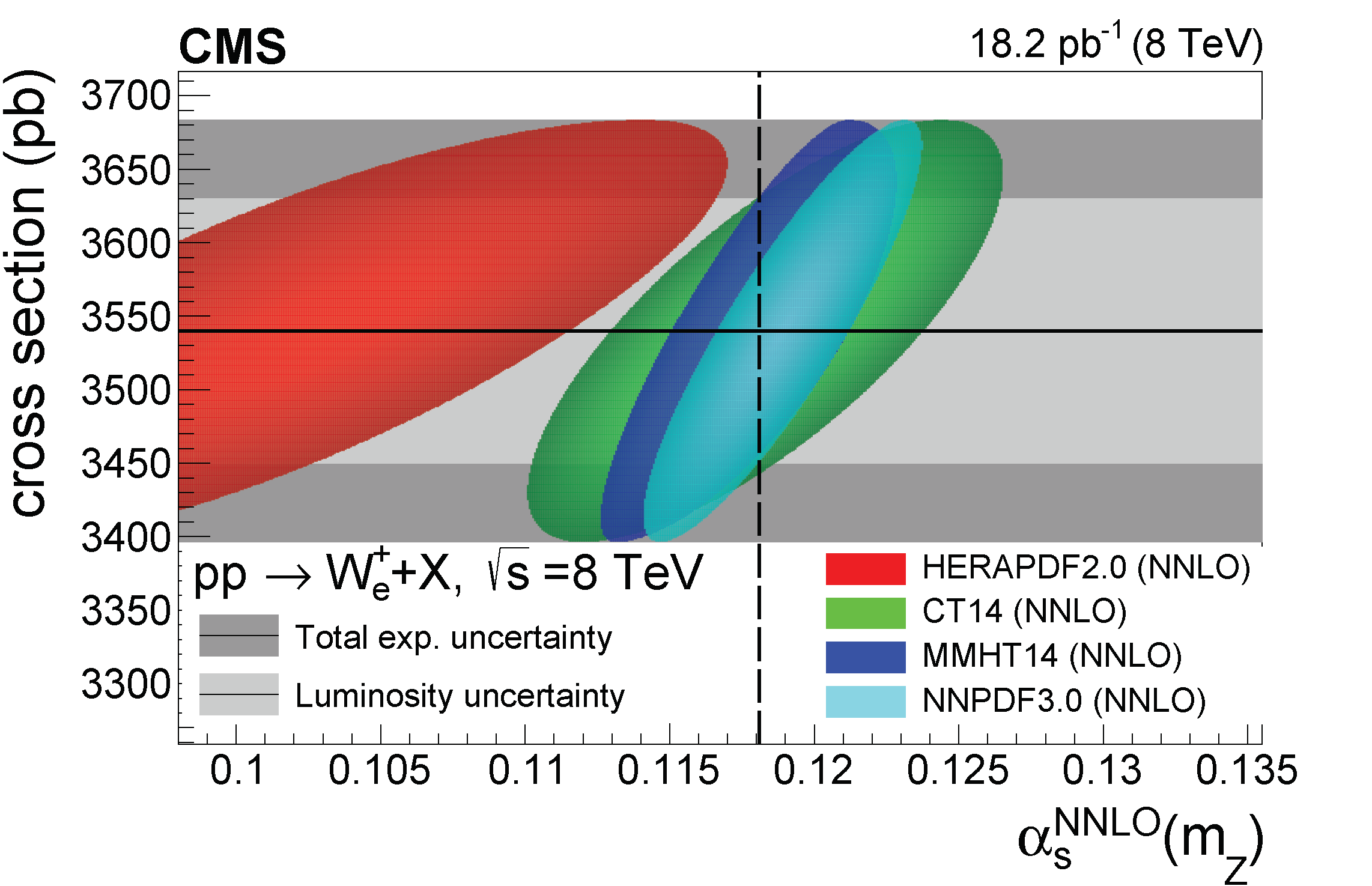}
\includegraphics[width=0.49\textwidth]{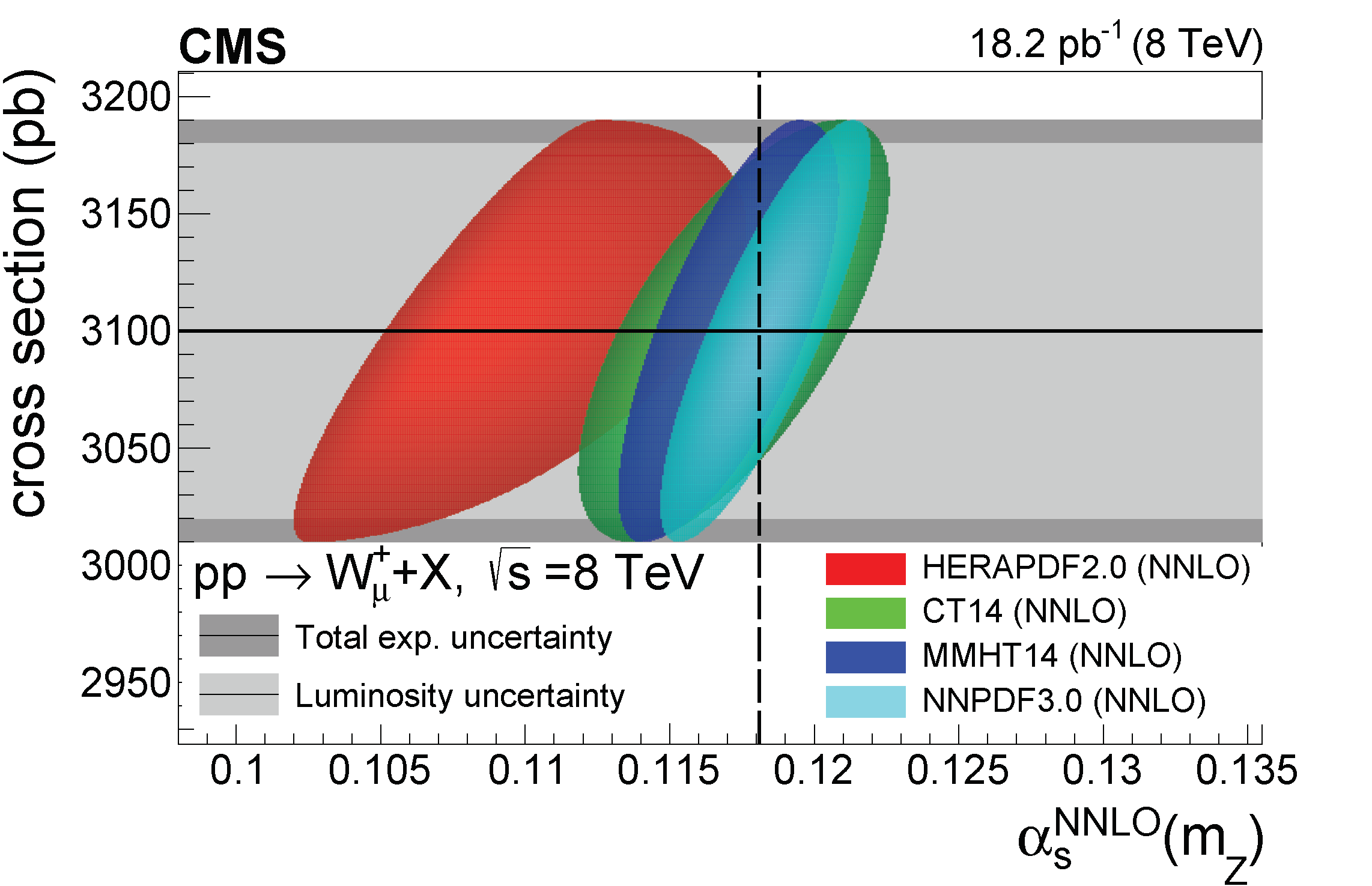}
\includegraphics[width=0.49\textwidth]{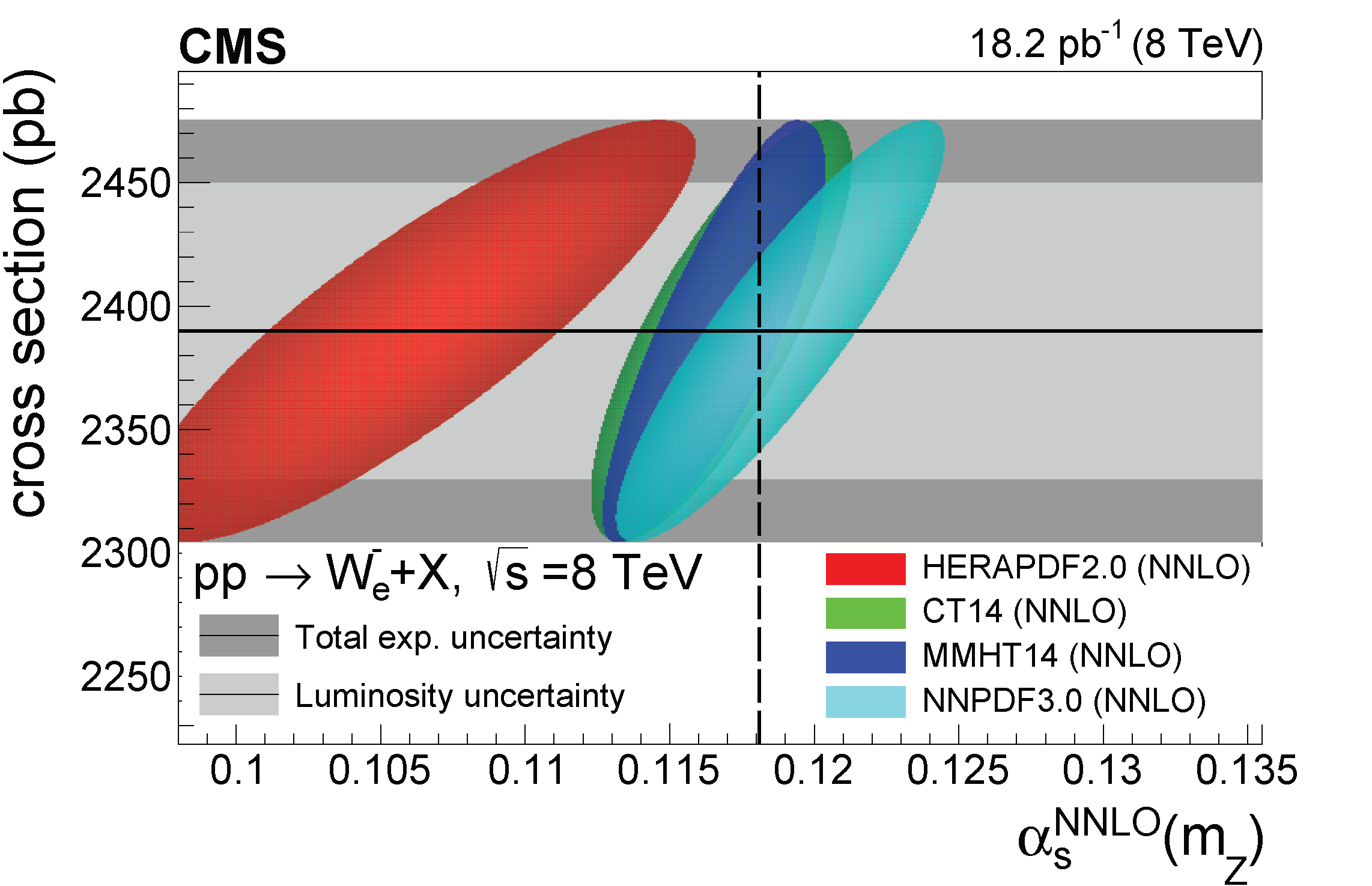}
\includegraphics[width=0.49\textwidth]{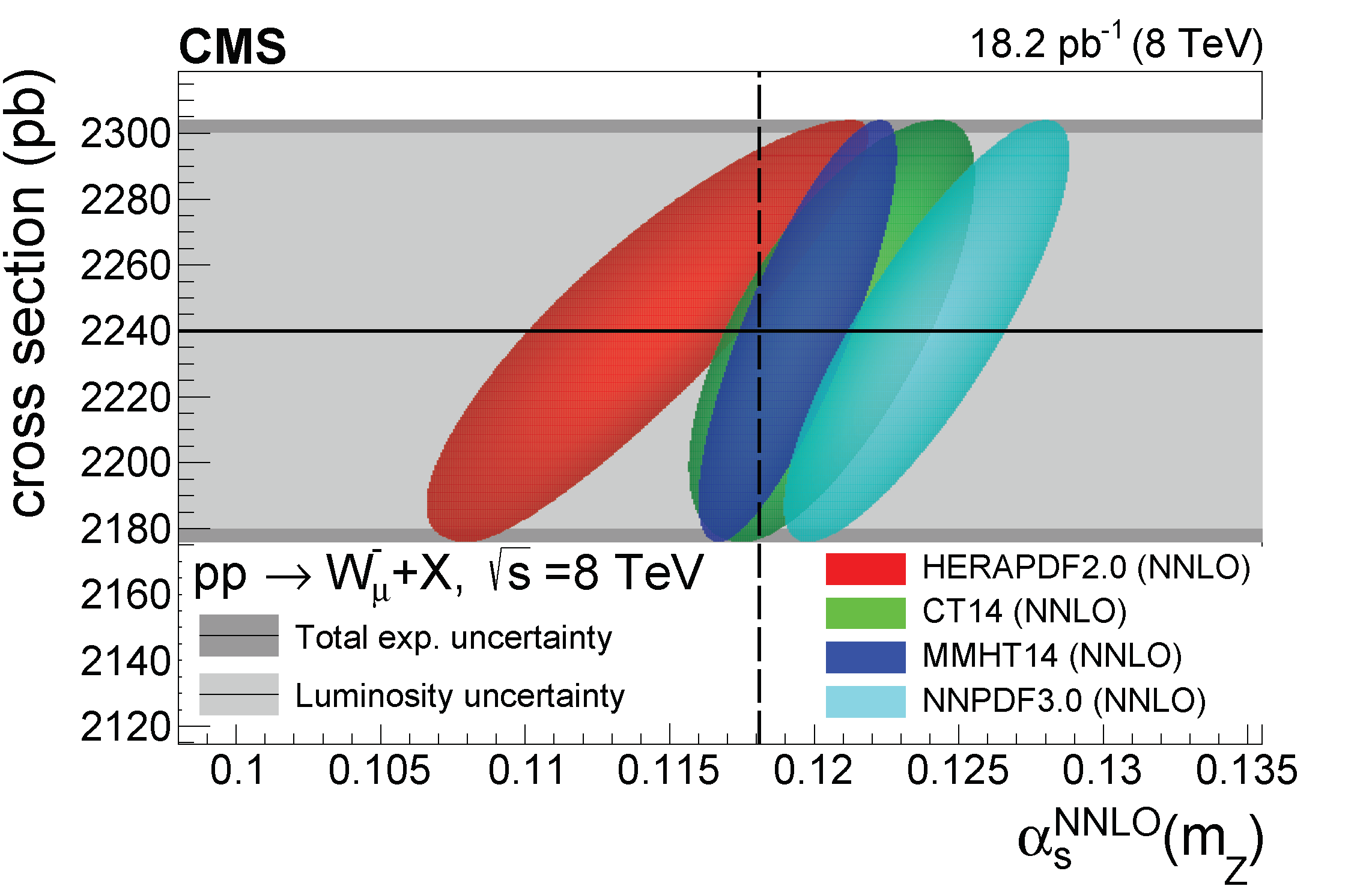}
\includegraphics[width=0.49\textwidth]{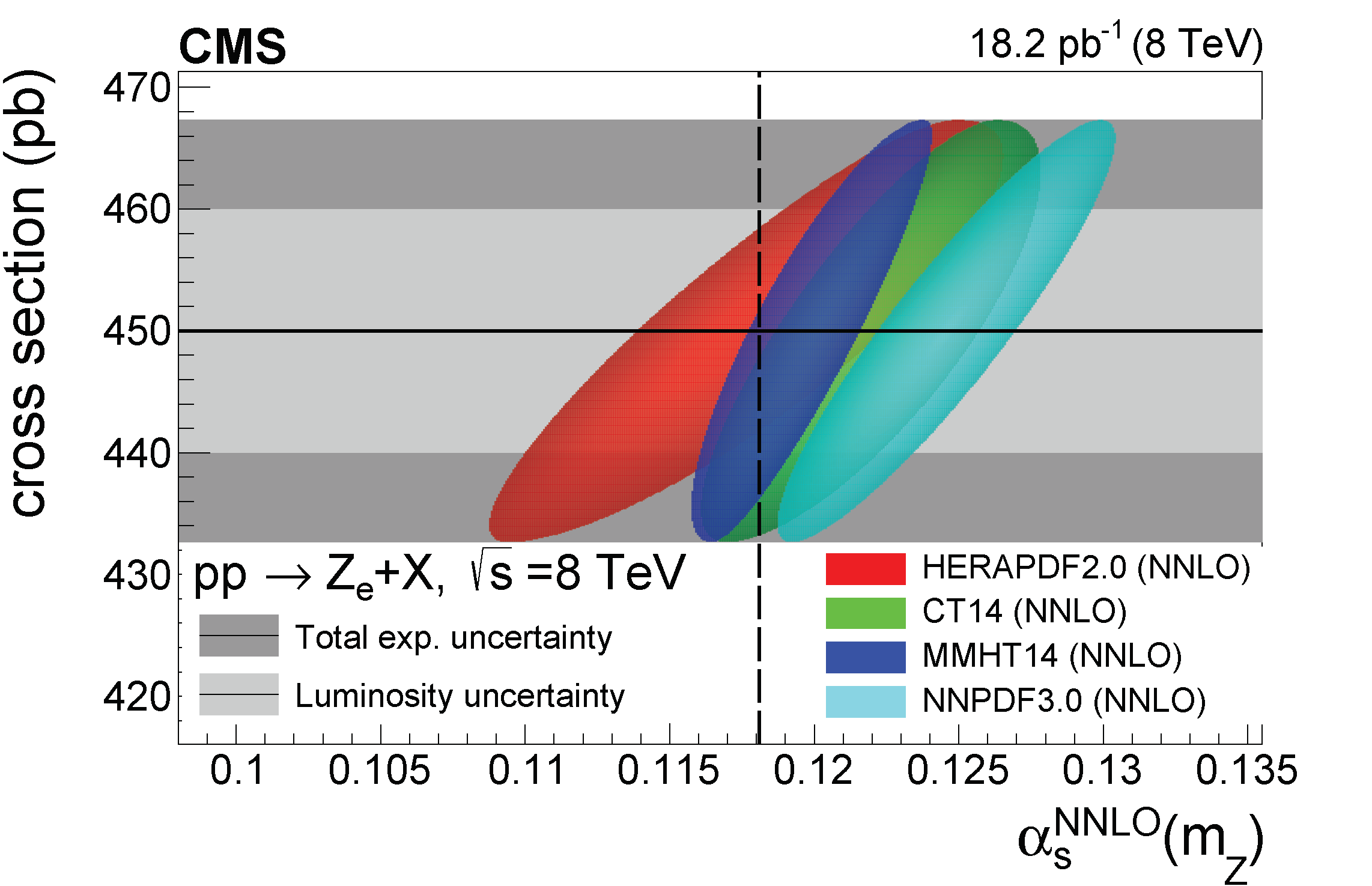}
\includegraphics[width=0.49\textwidth]{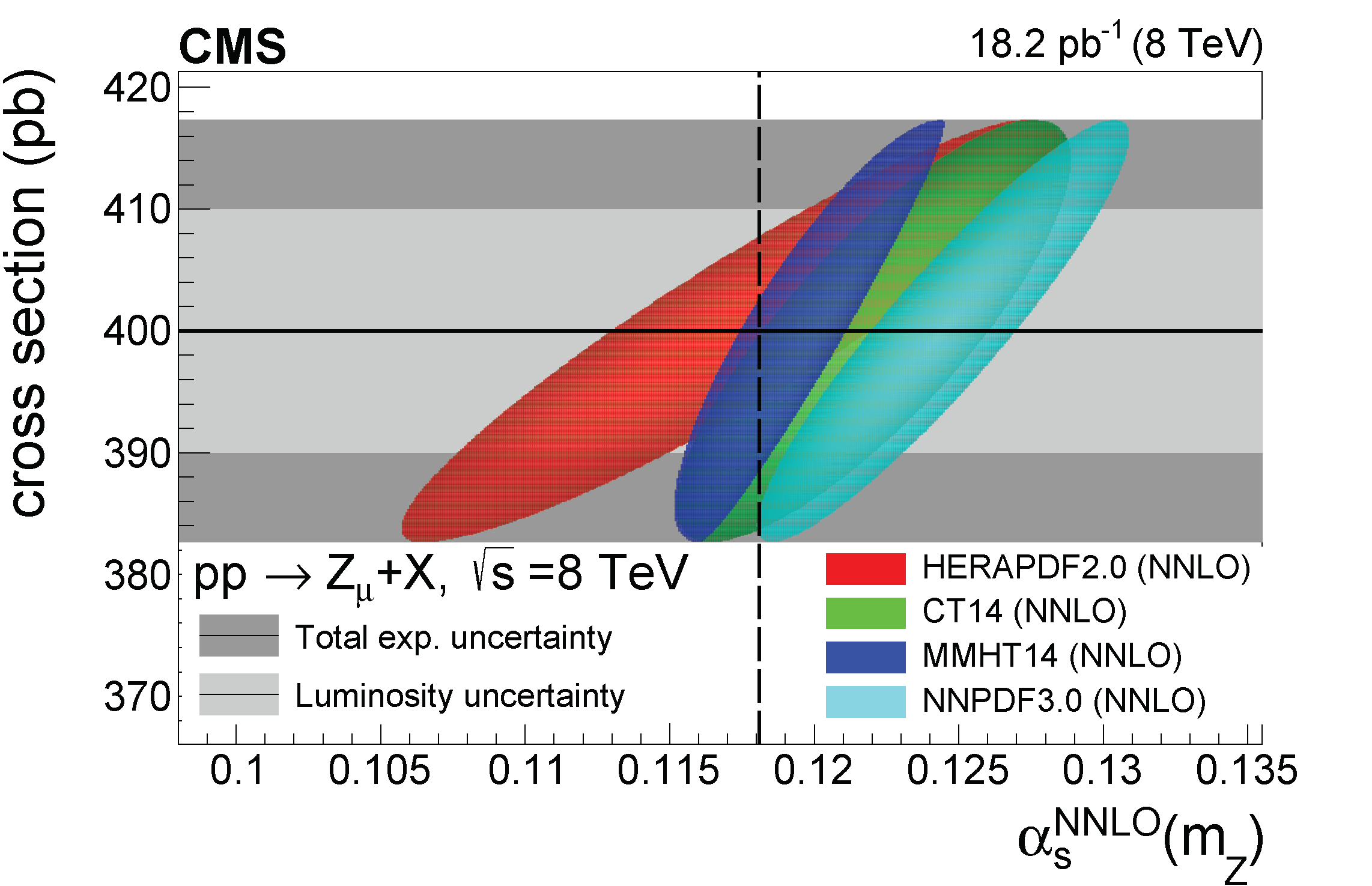}
\caption{Same as Fig.~\ref{fig:fitgraphfirst} for the production of $\PW^{+}_{\Pe}$ (upper left) and $\PW^{+}_{\PGm}$ (upper right),
$\PW^{-}_{\Pe}$ (middle left) and $\PW^{-}_{\PGm}$ (middle right), and $\PZ_{\Pe}$ (lower left) and $\PZ_{\PGm}$ (lower right) in $\Pp\Pp$ collisions at $\sqrts = 8\TeV$
\label{fig:fitgraphlast}.}
\end{figure}

For each of the twelve experimental \PWpm and \PZ boson cross section measurements listed in Table~\ref{tab:data}, we have computed
their corresponding theoretical NNLO pQCD predictions using the four PDF sets and five to seven different values
of $\alpS(m_{\PZ})$. It is important to stress again that, for each QCD coupling constant, we use the specific PDF sets that 
are associated with that particular $\alpS(m_{\PZ})$ value. We calculated the NLO EW corrections using NNPDF3.0 and $\alpS(m_{\PZ})=0.118$.
By comparing the whole set of theoretical calculations to the experimental data, a preferred value
of the QCD coupling constant can be derived for each PDF set as explained in the next section.
Figures~\ref{fig:fitgraphfirst}--\ref{fig:fitgraphlast} show the fiducial cross sections as a function of $\alpS(m_{\PZ})$, with the
experimental values indicated by the horizontal black line with the inner grey band showing the integrated luminosity uncertainty,
and the outer darker band showing the total experimental uncertainties added in quadrature. The filled ellipses represent the contours
of the joint probability density functions (Jpdfs) of the theoretical and experimental results, with a width representing a two-dimensional one standard deviation
obtained from the product of the probability densities of the experimental and theoretical results for each PDF, as
described in the next section.
For any fixed value of $\alpS(m_{\PZ})$, a hierarchy of \PWpm, \PZ theoretical cross sections is apparent with HERAPDF2.0 predictions tending to be
systematically above the data, and the NNPDF3.0 ones  below the latter. In between the results of these two PDF sets, the cross sections derived
with MMHT14 tend to be above those with CT14, although they are often very similar and overlap most of the time. Alternatively, the results
of Figs.~\ref{fig:fitgraphfirst}--\ref{fig:fitgraphlast} indicate that, in order to reproduce the experimental cross sections,
HERAPDF2.0 (NNPDF3.0) tends in general to prefer a smaller (larger) value of $\alpS(m_{\PZ})$ than other PDFs, and that the predictions
from CT14 and MMHT14 tend to be less scattered over the $\alpS(m_{\PZ})$ axis than those from HERAPDF2.0 and NNPDF3.0.
The HERAPDF2.0 (MMHT14) filled ellipses have the smallest (largest) relative slope as a function of $\alpS(m_{\PZ})$.
A larger slope is advantageous for extracting the strong coupling constant, because this means that the underlying $\alpS(m_{\PZ})$ value in the calculations has a
larger impact on the computed cross sections, also leading to a lower propagated uncertainty in the $\alpS(m_{\PZ})$ value derived by comparing
the theoretical prediction to the experimental value.

Overall, the theoretical predictions computed using the world-average value of the QCD coupling constant (vertical dashed line in
Figs.~\ref{fig:fitgraphfirst}--\ref{fig:fitgraphlast}) agree well with the experimental values within the uncertainties.
The level of data-theory agreement can be quantified with a goodness-of-fit test, $\chi^2 = \xi_i (M^{-1})_{ij} \xi_j$,
where $M$ is the covariance matrix taking all the uncertainties and their correlations into account,
as explained in Section~\ref{sec:combiningEstimates}, and $\xi_i=\sigma_{i,\text{th}}-\sigma_{i,\text{exp}}$ is the difference between theoretical and experimental
cross sections for each PDF set. In the $\chi^2$ calculation, the asymmetric uncertainties of the CT14, HERAPDF2.0, and MMHT14 PDF sets are symmetrised to the largest of
the two values and also separately to the smallest of the two values. The results are listed in Table~\ref{tab:chi2}.

\begin{table}[htpb!]
\topcaption{Overall goodness-of-fit per number of degrees of freedom, $\chi^2$/ndf, among the twelve experimental measurements of \PWpm
and \PZ boson production cross sections
and the corresponding theoretical calculations obtained with the four different PDF sets for their default $\alpS(m_{\PZ})=0.118$ value.
The first (second) row is obtained symmetrising the PDF uncertainties of the cross sections obtained with the CT14, HERAPDF2.0, and MMHT14 sets
to the largest (smallest) of their respective values.\label{tab:chi2}}
\centering
\cmsTable{
\begin{tabular}{lllll}\\\hline
& CT14 & HERAPDF2.0 & MMHT14 & NNPDF3.0 \\
$\chi^2/$ndf (symmetrised to the largest PDF uncertainty value) & 12.0/11 & 11.0/11 & 9.0/11 & 31.7/11 \\
$\chi^2/$ndf (symmetrised to the smallest PDF uncertainty value) & 15.2/11 & 28.3/11 & 13.9/11 & 31.7/11 \\\hline
\end{tabular}
}
\end{table}

\section{Extraction of the QCD coupling constant}
\label{sec:4}

\subsection{Extraction of \texorpdfstring{$\alpS(m_{\PZ})$}{alpha\_S} for each single \texorpdfstring{\PWpm}{W+-} and \texorpdfstring{\PZ}{Z} cross section measurement}
\label{sec:extract_one_estimate}

The dependence of the theoretical cross sections on the QCD coupling constant $\alpS(m_{\PZ})$,
shown in Figs.~\ref{fig:fitgraphfirst}--\ref{fig:fitgraphlast}, is fitted through a linear $\chi^2$-minimisation procedure
over $\alpS(m_{\PZ})\in[0.115,0.121]$, to extract the slope $k$.
Over the considered $\alpS(m_{\PZ})$ range, the empirical linear fit describes well the observed $\alpS$-dependence of the theoretical cross section for all PDF sets.
The value of $\alpS(m_{\PZ})$ preferred by each individual measurement is determined by the crossing point of the fitted
linear theoretical curve with the experimental horizontal line. The resulting $\alpS(m_{\PZ})$ values are listed in Table~\ref{tab:all_alphas}.
For each theoretical point used in the fit, the uncertainty in the cross section
is given by the quadratic sum of its associated PDF, scale, and numerical uncertainties.
In order to exploit the dependence of $\sigma_\text{th}(\PV)$ on $\alpS$ to quantitatively derive the latter,
a joint probability density function is constructed for each PDF prediction, as explained next.
When the theoretical cross section value has positive and negative uncertainties $\delta_\text{+-}$,
the experimental cross section is $\sigma_\text{exp}$ with uncertainty $\delta_\text{exp}$, the slope of the fit is $k$, and the fitted QCD
coupling constant is $\alpS(m_{\PZ})$, then the Jpdf as a function of $\sigma$ and $\alpS$ is proportional to
\begin{linenomath*}
\begin{equation}\label{eq:ellipsoids}
\exp\left\{-\frac12
	\left[
	\left(\frac{\sigma-\sigma_\text{exp}}{\delta_\text{exp}}\right)^2 +
	\left(\frac{\sigma-\sigma_\text{exp}-k (\alpS-\alpS(m_{\PZ}))}{\delta_\text{+-}}\right)^2
	\right]\right\}.
\end{equation}
\end{linenomath*}
The sign of $(\sigma-\sigma_\text{exp}-k (\alpS-\alpS(m_{\PZ})))$ determines which of the $\delta_\text{+-}$ is used.
For symmetric uncertainties the Jpdfs have elliptical contours, but for asymmetric ones they are two filled ellipses combined together.
This procedure is repeated for all the twelve different measurements and for all four PDF sets, and plotted as the filled ellipses
shown in Figs.~\ref{fig:fitgraphfirst}--\ref{fig:fitgraphlast}, where each coloured area corresponds to one two-dimensional
(in $\sigma$ and $\alpS(m_{\PZ})$) standard deviation.

\begin{table}[htbp!]
\topcaption{Extracted $\alpS(m_{\PZ})$ values from the different data-theory \PWpm and \PZ boson production cross section
comparisons for each PDF set, with associated uncertainties from different experimental
(statistical, integrated luminosity, and systematic) and theoretical (PDF, scale, and numerical) sources.}
\centering
\cmsTable{
\renewcommand*{\arraystretch}{1.2}
\begin{tabular}{llccccccc}\hline
Cross section & PDF & $\alpS(m_{\PZ})$
(total) & $\delta_{\alpS}\stat$ & $\delta_{\alpS}\lum$ & $\delta_{\alpS}\syst$ & $\delta_{\alpS}$(PDF) & $\delta_{\alpS}$(scale) & $\delta_{\alpS}\,\statt$ \\\hline
$\PW^{+}_{\Pe}$ (7\TeV) & CT14 & 0.1193 $^{+0.0062}_{-0.0062}$ & 0.0004 & 0.0046 & 0.0022 & $^{+0.0031}_{-0.0032}$ & 0.0017 & 0.0006 \\
& HERAPDF2.0 & 0.1108 $^{+0.0090}_{-0.0097}$ & 0.0006 & 0.0072 & 0.0035 & $^{+0.0033}_{-0.0050}$ & 0.0017 & 0.0012 \\
& MMHT14 & 0.1179 $^{+0.0049}_{-0.0047}$ & 0.0003 & 0.0037 & 0.0018 & $^{+0.0025}_{-0.0020}$ & 0.0008 & 0.0005 \\
& NNPDF3.0 & 0.1200 $^{+0.0054}_{-0.0054}$ & 0.0004 & 0.0043 & 0.0021 & $^{+0.0022}_{-0.0022}$ & 0.0009 & 0.0006 \\

$\PW^{-}_{\Pe}$ (7\TeV) & CT14 & 0.1208 $^{+0.0064}_{-0.0061}$ & 0.0005 & 0.0047 & 0.0022 & $^{+0.0034}_{-0.0030}$ & 0.001 & 0.0004 \\
& HERAPDF2.0 & 0.1152 $^{+0.0136}_{-0.0149}$ & 0.0013 & 0.0118 & 0.0056 & $^{+0.0027}_{-0.0066}$ & 0.0025 & 0.0009 \\
& MMHT14 & 0.1195 $^{+0.0047}_{-0.0053}$ & 0.0004 & 0.0040 & 0.0019 & $^{+0.0012}_{-0.0027}$ & 0.0007 & 0.0003 \\
& NNPDF3.0 & 0.1239 $^{+0.0073}_{-0.0073}$ & 0.0007 & 0.0060 & 0.0028 & $^{+0.0029}_{-0.0029}$ & 0.0011 & 0.0005 \\

$\PZ_{\Pe}$ (7\TeV) & CT14 & 0.1247 $^{+0.0068}_{-0.0070}$ & 0.0014 & 0.0051 & 0.0028 & $^{+0.0031}_{-0.0036}$ & 0.0004 & 0.0003 \\
& HERAPDF2.0 & 0.1226 $^{+0.0106}_{-0.0121}$ & 0.0025 & 0.0088 & 0.0049 & $^{+0.0018}_{-0.0061}$ & 0.0007 & 0.0005 \\
& MMHT14 & 0.1222 $^{+0.0047}_{-0.0050}$ & 0.0011 & 0.0038 & 0.0021 & $^{+0.0012}_{-0.0022}$ & 0.0006 & 0.0002 \\
& NNPDF3.0 & 0.1279 $^{+0.0074}_{-0.0074}$ & 0.0016 & 0.0058 & 0.0032 & $^{+0.0027}_{-0.0027}$ & 0.0007 & 0.0003 \\

$\PW^{+}_{\PGm}$ (7\TeV) & CT14 & 0.1178 $^{+0.0049}_{-0.0058}$ & 0.0003 & 0.0040 & 0.0015 & $^{+0.0023}_{-0.0039}$ & 0.0007 & 0.0005 \\
& HERAPDF2.0 & 0.1085 $^{+0.0083}_{-0.0108}$ & 0.0006 & 0.0070 & 0.0026 & $^{+0.0026}_{-0.0073}$ & 0.0023 & 0.0009 \\
& MMHT14 & 0.1170 $^{+0.0048}_{-0.0053}$ & 0.0003 & 0.0039 & 0.0015 & $^{+0.0022}_{-0.0031}$ & 0.0006 & 0.0006 \\
& NNPDF3.0 & 0.1185 $^{+0.0054}_{-0.0054}$ & 0.0003 & 0.0044 & 0.0016 & $^{+0.0022}_{-0.0022}$ & 0.0011 & 0.0006 \\

$\PW^{-}_{\PGm}$ (7\TeV) & CT14 & 0.1186 $^{+0.0050}_{-0.0057}$ & 0.0004 & 0.0041 & 0.0014 & $^{+0.0023}_{-0.0036}$ & 0.0009 & 0.0003 \\
& HERAPDF2.0 & 0.1109 $^{+0.0111}_{-0.0109}$ & 0.001 & 0.0094 & 0.0033 & $^{+0.0040}_{-0.0035}$ & 0.0023 & 0.0008 \\
& MMHT14 & 0.1177 $^{+0.0046}_{-0.0047}$ & 0.0004 & 0.0039 & 0.0014 & $^{+0.0017}_{-0.0021}$ & 0.0009 & 0.0003 \\
& NNPDF3.0 & 0.1212 $^{+0.0070}_{-0.0070}$ & 0.0006 & 0.0058 & 0.0020 & $^{+0.0029}_{-0.0029}$ & 0.0013 & 0.0004 \\

$\PZ_{\PGm}$ (7\TeV) & CT14 & 0.1232 $^{+0.0062}_{-0.0077}$ & 0.001 & 0.0051 & 0.0022 & $^{+0.0023}_{-0.0051}$ & 0.0005 & 0.0003 \\
& HERAPDF2.0 & 0.1200 $^{+0.0108}_{-0.0108}$ & 0.0017 & 0.0092 & 0.0040 & $^{+0.0034}_{-0.0034}$ & 0.0012 & 0.0005 \\
& MMHT14 & 0.1213 $^{+0.0051}_{-0.0045}$ & 0.0007 & 0.0039 & 0.0017 & $^{+0.0027}_{-0.001}$ & 0.0005 & 0.0002 \\
& NNPDF3.0 & 0.1261 $^{+0.0070}_{-0.0070}$ & 0.0011 & 0.0057 & 0.0025 & $^{+0.0028}_{-0.0028}$ & 0.0007 & 0.0003 \\[\cmsTabSkip]

$\PW^{+}_{\Pe}$ (8\TeV) & CT14 & 0.1181 $^{+0.0081}_{-0.0083}$ & 0.0009 & 0.0039 & 0.0047 & $^{+0.0049}_{-0.0053}$ & 0.0015 & 0.0009 \\
& HERAPDF2.0 & 0.1030 $^{+0.0154}_{-0.0140}$ & 0.0015 & 0.0070 & 0.0085 & $^{+0.0099}_{-0.0075}$ & 0.0037 & 0.0017 \\
& MMHT14 & 0.1172 $^{+0.0045}_{-0.0057}$ & 0.0006 & 0.0025 & 0.0031 & $^{+0.0017}_{-0.0039}$ & 0.0011 & 0.0006 \\
& NNPDF3.0 & 0.1188 $^{+0.0049}_{-0.0049}$ & 0.0006 & 0.0027 & 0.0032 & $^{+0.0022}_{-0.0022}$ & 0.0012 & 0.0006 \\

$\PW^{-}_{\Pe}$ (8\TeV) & CT14 & 0.1169 $^{+0.0046}_{-0.0044}$ & 0.0004 & 0.0025 & 0.0025 & $^{+0.0029}_{-0.0025}$ & 0.0006 & 0.0003 \\
& HERAPDF2.0 & 0.1066 $^{+0.0098}_{-0.0094}$ & 0.001 & 0.0057 & 0.0057 & $^{+0.0049}_{-0.0042}$ & 0.0020 & 0.0009 \\
& MMHT14 & 0.1163 $^{+0.0036}_{-0.0041}$ & 0.0004 & 0.0022 & 0.0022 & $^{+0.0014}_{-0.0025}$ & 0.001 & 0.0003 \\
& NNPDF3.0 & 0.1187 $^{+0.0059}_{-0.0059}$ & 0.0006 & 0.0035 & 0.0035 & $^{+0.0029}_{-0.0029}$ & 0.0008 & 0.0005 \\

$\PZ_{\Pe}$ (8\TeV) & CT14 & 0.1216 $^{+0.0056}_{-0.0062}$ & 0.0027 & 0.0027 & 0.0027 & $^{+0.0029}_{-0.0040}$ & 0.0007 & 0.0003 \\
& HERAPDF2.0 & 0.1173 $^{+0.0084}_{-0.0093}$ & 0.0045 & 0.0045 & 0.0045 & $^{+0.0031}_{-0.0051}$ & 0.0009 & 0.0005 \\
& MMHT14 & 0.1201 $^{+0.0044}_{-0.0039}$ & 0.0021 & 0.0021 & 0.0021 & $^{+0.0023}_{-0.0013}$ & 0.0006 & 0.0002 \\
& NNPDF3.0 & 0.1245 $^{+0.0060}_{-0.0060}$ & 0.0031 & 0.0031 & 0.0031 & $^{+0.0027}_{-0.0027}$ & 0.0006 & 0.0003 \\

$\PW^{+}_{\PGm}$ (8\TeV) & CT14 & 0.1173 $^{+0.0055}_{-0.0053}$ & 0.0004 & 0.0032 & 0.0016 & $^{+0.0038}_{-0.0035}$ & 0.0014 & 0.0008 \\
& HERAPDF2.0 & 0.1076 $^{+0.0055}_{-0.0101}$ & 0.0006 & 0.0045 & 0.0022 & $^{+0.0011}_{-0.0085}$ & 0.0019 & 0.0009 \\
& MMHT14 & 0.1168 $^{+0.0035}_{-0.0041}$ & 0.0003 & 0.0024 & 0.0012 & $^{+0.0020}_{-0.0029}$ & 0.0009 & 0.0005 \\
& NNPDF3.0 & 0.1182 $^{+0.0038}_{-0.0038}$ & 0.0003 & 0.0027 & 0.0013 & $^{+0.0022}_{-0.0022}$ & 0.0007 & 0.0006 \\

$\PW^{-}_{\PGm}$ (8\TeV) & CT14 & 0.1209 $^{+0.0053}_{-0.0046}$ & 0.0005 & 0.0032 & 0.0011 & $^{+0.0039}_{-0.0030}$ & 0.0008 & 0.0003 \\
& HERAPDF2.0 & 0.1147 $^{+0.0081}_{-0.0072}$ & 0.0010 & 0.0062 & 0.0021 & $^{+0.0041}_{-0.0021}$ & 0.0018 & 0.0007 \\
& MMHT14 & 0.1195 $^{+0.0035}_{-0.0034}$ & 0.0004 & 0.0026 & 0.0009 & $^{+0.0019}_{-0.0016}$ & 0.0009 & 0.0003 \\
& NNPDF3.0 & 0.1238 $^{+0.0052}_{-0.0052}$ & 0.0006 & 0.0039 & 0.0013 & $^{+0.0029}_{-0.0029}$ & 0.0011 & 0.0004 \\

$\PZ_{\PGm}$ (8\TeV) & CT14 & 0.1220 $^{+0.0067}_{-0.0069}$ & 0.0032 & 0.0032 & 0.0032 & $^{+0.0037}_{-0.0040}$ & 0.0006 & 0.0003 \\
& HERAPDF2.0 & 0.1170 $^{+0.0112}_{-0.0116}$ & 0.0060 & 0.0060 & 0.0060 & $^{+0.0036}_{-0.0048}$ & 0.0013 & 0.0006 \\
& MMHT14 & 0.1202 $^{+0.0050}_{-0.0043}$ & 0.0024 & 0.0024 & 0.0024 & $^{+0.0027}_{-0.0007}$ & 0.0005 & 0.0002 \\
& NNPDF3.0 & 0.1244 $^{+0.0066}_{-0.0066}$ & 0.0034 & 0.0034 & 0.0034 & $^{+0.0028}_{-0.0028}$ & 0.0006 & 0.0003 \\ \hline
\end{tabular}
}
\label{tab:all_alphas}
\end{table}

\subsection{Propagation of \texorpdfstring{$\alpS(m_{\PZ})$}{alpha\_S} uncertainties}
\label{sec:uncert_propag}

Appropriate propagation of the separated experimental and theoretical uncertainties into each value of
$\alpS(m_{\PZ})$ obtained from each particular \PWpm and \PZ measurement, is crucial to combine all estimates
taking into account their correlations, and extract a single final $\alpS(m_{\PZ})$ result.
The method employed to determine the individual sources of uncertainties associated with a given $\alpS(m_{\PZ})$ value
is similar to that used in Refs.~\cite{Chatrchyan:2013haa,Klijnsma:2017eqp} for the $\alpS(m_{\PZ})$ extraction from inclusive
$\ttbar$ cross sections. In summary, each source of uncertainty $\delta\sigma$ propagates into a corresponding $\alpS(m_{\PZ})$ uncertainty
through $\delta\sigma/k$, where $k$ is the slope of the fit of the theoretical cross section versus $\alpS(m_{\PZ})$.
The validity of such a simple propagation of uncertainties can be demonstrated calculating first a theoretical uncertainty distribution by
adding up quadratically the PDF, scale, and numerical uncertainties assuming Gaussian distributions (for the asymmetric PDF uncertainties,
only the largest of the positive and negative uncertainties are used). Then, a Jpdf can be derived from
the product of the theoretical and experimental distributions
$f_{\text{th}}(\sigma | \alpS(m_{\PZ}))$ and $f_{\text{exp}}(\sigma)$. Integration over $\sigma$ gives the marginalised posterior
\begin{linenomath*}
\begin{equation*}
P(\alpS) = \int f_{\text{exp}}(\sigma)\, f_{\text{th}}(\sigma | \alpS) \,\rd\sigma.
\end{equation*}
\end{linenomath*}
The expected value of the theoretical probability distribution changes linearly according to the fitted first-order polynomial,
but all the theoretical and experimental uncertainties remain the same for all $\alpS(m_{\PZ})$ values. Since all the uncertainties
have Gaussian distributions and the marginalisation is, in essence, a convolution, the resulting $\alpS(m_{\PZ})$ posterior will be
also Gaussian, with the impact of each cross section uncertainty adding quadratically to the $\alpS(m_{\PZ})$ uncertainty.
More specifically, each cross section uncertainty source $\delta\sigma$ will result in a propagated $\alpS(m_{\PZ})$ uncertainty in
$\delta\sigma/k$ size, where $k$ is the slope of the linear fit to theoretical calculations. In this demonstration, we symmetrised
the PDF uncertainties for simplicity, but the $\delta\sigma/k$ prescription will be also used hereafter for the case of
asymmetric uncertainties. All the extracted $\alpS(m_{\PZ})$ values,
along with the uncertainty breakdowns from every source, for each system and PDF set are given in Table~\ref{tab:all_alphas}.
The results from the MMHT14 PDF feature the extracted $\alpS(m_{\PZ})$ values with the lowest overall uncertainty, in some cases as low as 3\%.

\subsection{Combination of all individual \texorpdfstring{$\alpS(m_{\PZ})$}{alpha\_S} values}
\label{sec:combiningEstimates}

From the twelve $\alpS(m_{\PZ})$ extractions per PDF set listed in Table~\ref{tab:all_alphas}, we can determine a single $\alpS(m_{\PZ})$ value by
appropriately combining them taking into account their uncorrelated, partially-, and fully-correlated experimental and theoretical uncertainties.
For this, the program \textsc{convino} v1.2~\cite{Kieseler:2017kxl} is employed, which uses a $\chi^2$ minimisation to determine the best estimate.
In the current analysis, the Neyman $\chi^2$ code option is always used.
As an independent cross-check, we confirm that, for symmetric uncertainties, the results are identical to those obtained with
the BLUE method~\cite{Nisius:2014wua}. The following correlation coefficients are used:
\begin{itemize}
\item The integrated luminosity uncertainty is fully correlated for all $\alpS(m_{\PZ})$ results obtained at the same $\sqrts$,
but fully uncorrelated between the two different c.m.\ energies.
\item The experimental systematic uncertainty is partially correlated. Since the exact correlation values impact the final
result, a dedicated study of the correlations is carried out in Section~\ref{sec:expsystcorr}.
\item The experimental statistical uncertainty is fully uncorrelated among $\alpS(m_{\PZ})$ extractions.
\item The PDF uncertainty is partially correlated for the $\alpS(m_{\PZ})$ values extracted with the same PDF set, as discussed in detail
in Section~\ref{sec:pdfscalecorr}.
\item The scale uncertainty is partially correlated, as explained in Section~\ref{sec:pdfscalecorr}.
\item The theoretical numerical uncertainty is fully uncorrelated among $\alpS(m_{\PZ})$ extractions.
\end{itemize}
By properly implementing all the uncertainties and their correlations in \textsc{convino}, we can derive a single final combined
$\alpS(m_{\PZ})$ value and associated uncertainties for each PDF set.

\subsubsection{Correlations among the experimental systematic uncertainties
\label{sec:expsystcorr}}

For all the experimental measurements of the cross sections, the size of their systematic uncertainties of each type are
listed in Table~\ref{tab:expuncerts}. The absolute uncertainties are given in the same proportions as in Table~\ref{tab:expuncerts},
but rescaled such that they add up quadratically to the total experimental systematic uncertainty listed in Tables~\ref{tab:CMS7_x} and \ref{tab:CMS8_x}.

\begin{table}[!hbt]
\topcaption{Breakdown of the experimental systematic uncertainties (in percent) for each of the \PWpm and
\PZ boson production cross section measurements at 7 and 8\TeV~\cite{CMS:2011aa,Chatrchyan:2014mua}.
\label{tab:expuncerts}}
\centering
\begin{tabular}{lcccccc}\hline
Measurement & $\PW^{+}_{\Pe}$ & $\PW^{-}_{\Pe}$ & $\PZ_{\Pe}$ & $\PW^{+}_{\PGm}$ & $\PW^{-}_{\PGm}$ & $\PZ_{\PGm}$ \\\hline
7\TeV &  &  &  &  &  &  \\
Lepton reconstruction and identification & 1.5 & 1.5 & 1.8 & 0.9 & 0.9 & \NA \\
Muon trigger inefficiency & \NA & \NA & \NA & 0.5 & 0.5 & 0.5 \\
Energy scale and resolution & 0.5 & 0.6 & 0.12 & 0.19 & 0.25 & 0.35 \\
Missing $\pt$ scale and resolution & 0.3 & 0.3 & \NA & 0.2 & 0.2 & \NA \\
Background subtraction and modelling & 0.3 & 0.5 & 0.14 & 0.4 & 0.5 & 0.28 \\[\cmsTabSkip]
8\TeV &  &  &  &  &  &  \\
Lepton reconstruction and identification & 2.8 & 2.5 & 2.8 & 1.0 & 0.9 & 1.1 \\
Energy scale and resolution & 0.4 & 0.7 & 0.0 & 0.3 & 0.3 & 0.0 \\
Missing $\pt$ scale and resolution & 0.8 & 0.7 & \NA & 0.5 & 0.5 & \NA \\
Background subtraction and modelling & 0.2 & 0.3 & 0.4 & 0.2 & 0.1 & 0.4  \\ \hline
\end{tabular}
\end{table}

The detailed correlations between the different uncertainty sources of Table~\ref{tab:expuncerts} are listed in
Tables~\ref{tab:expcorrfirst} to \ref{tab:expcorrlast} of the Appendix. By using the experimental systematic
uncertainty breakdown and the correlations between the uncertainties, the total correlations between the
experimental uncertainty sources can be calculated using the formula
\begin{linenomath*}
\begin{equation}
\rho_{ij} = \frac{\sum_k \rho_{k,ij}  \,\delta\sigma_{k,i} \,\delta\sigma_{k,j}}{\sqrt{\sum_k \delta\sigma_{k,i}^2} \, \sqrt{\sum_k \delta\sigma_{k,j}^2}},
\end{equation}
\end{linenomath*}
where the subscript $k$ labels the uncertainty (\eg background subtraction), and $i$, $j$ denote the associated measurement
(\eg $\PW^{+}_{\Pe}$ at 7\TeV). The calculated total correlations among experimental systematic uncertainties
are given in Table~\ref{tab:exptotalcorr} of the Appendix. Many of the propagated experimental uncertainties appear
strongly correlated.
To give an idea of the correlations among $\alpS(m_{\PZ})$ estimates taking into account all the uncertainty sources, both
experimental and theoretical, an example correlation matrix for the NNPDF3.0 set is given in Table~\ref{tab:totalcorr1}
of the Appendix. One can see that many of the $\alpS(m_{\PZ})$ values derived within a given PDF set are strongly correlated, especially
across the same $\sqrts$.

\subsubsection{Correlations among  PDF and scale uncertainties
\label{sec:pdfscalecorr}}

In the theoretical cross section calculations, the PDF uncertainties are in the range of a few percent, scale uncertainty up to one percent,
and numerical uncertainty around half a percent (Tables~\ref{tab:CMS7_x} and \ref{tab:CMS8_x}).
The \MCFM numerical uncertainty cannot be neglected because differences with respect
to the prediction computed with the central eigenvalue/replica members are used to calculate the PDF uncertainties. The cross sections for
the central PDF members have intrinsic numerical fluctuations that impact the PDF uncertainty magnitude, asymmetry,
and also the correlations among theoretical uncertainties.
To take this into account, a Pearson correlation coefficient is calculated for each pair of measurements and for each PDF set
using the cross sections from all the PDF members that were used in calculating the PDF uncertainty. Similarly for the scale uncertainties,
for each pair of measurements the Pearson correlation coefficient is calculated using the results obtained from varying the theoretical scales.
The correlations are mostly in the 0.8--0.9, 0.4--0.7, 0.2--0.6, and 0.9--1.0 ranges for CT14, HERAPDF2.0, MMHT14, and
NNPDF3.0, respectively. The scale correlations are around 0.6--0.9. When combining the $\alpS(m_{\PZ})$
estimates, the specific correlation coefficient calculated for every specific pair of estimates is used.

\section{Results and discussion}
\label{sec:5}

\begin{figure}[htbp!]
\centering
\includegraphics[width=0.9\textwidth]{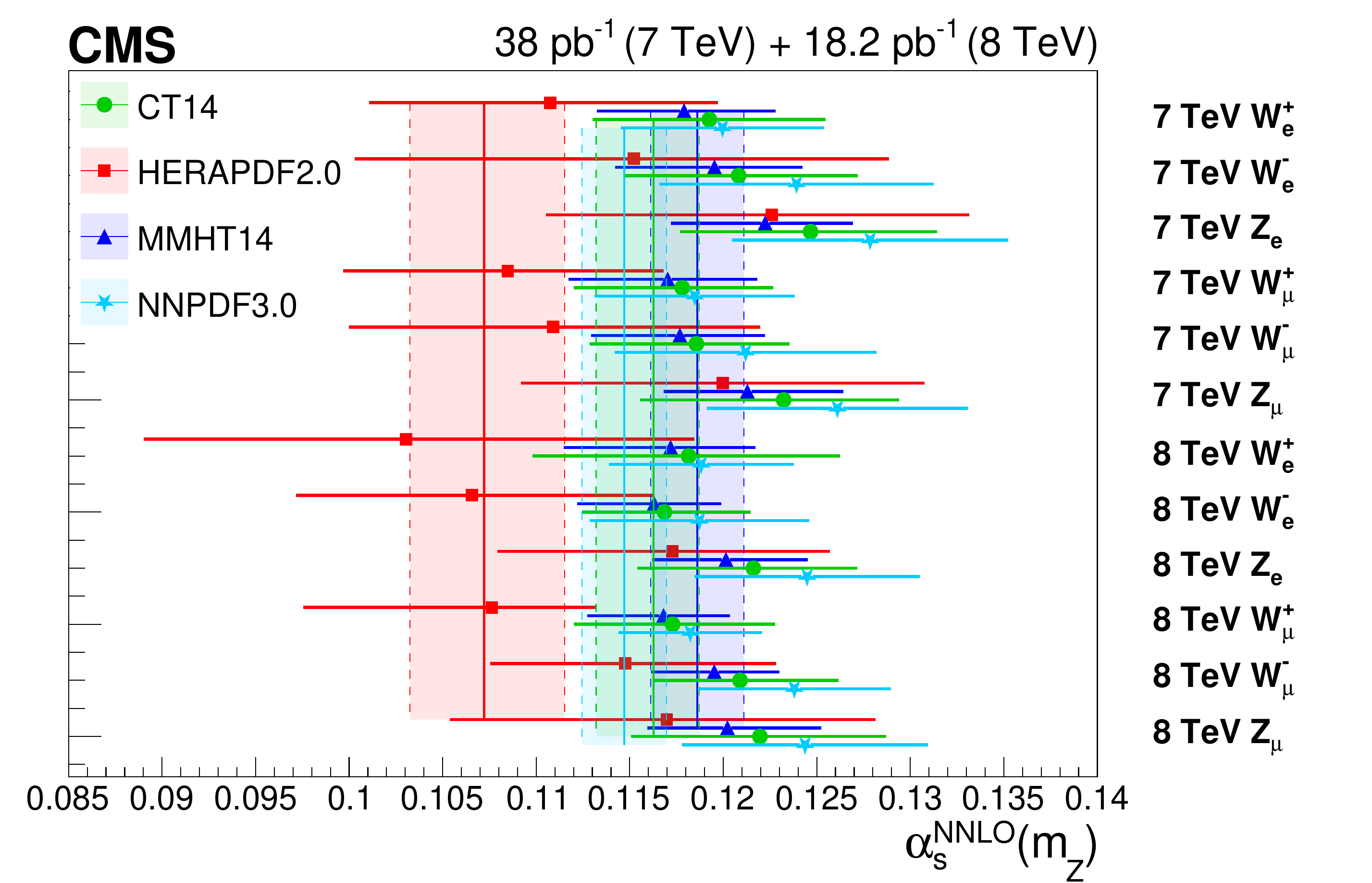}
\caption{Individual $\alpS(m_{\PZ})$ values extracted from each measured \PWpm and \PZ boson production cross section (bars), and final $\alpS(m_{\PZ})$
values obtained combining the twelve individual determinations (vertical coloured areas), for each PDF set.
\label{fig:allests}}
\end{figure}

Figure~\ref{fig:allests} shows the individual results (error bars) and the final combined $\alpS(m_{\PZ})$ value (coloured areas) obtained per PDF,
as explained in the previous section.
The width of the coloured areas in the plot indicates the size of the total propagated uncertainty in the final $\alpS(m_{\PZ})$ derived for each PDF set.
Table~\ref{tab:final_alphas_perPDF} lists the final $\alpS(m_{\PZ})$ values determined for each PDF set through the combination of the twelve
individual extractions. The total uncertainties amount to 2.0\% for MMHT14 and NNPDF3.0,
2.3\% for CT14, and $\approx$4\% for HERAPDF2.0. The total $\alpS(m_{\PZ})$ uncertainties derived for NNPDF3.0 are symmetric by
construction, and are also symmetric at the end for MMHT14 within the accuracy given. Small asymmetries remain for the
final CT14 and HERAPDF2.0 results.
The dominant source of experimental uncertainty is the integrated luminosity, whereas on the theoretical side it is the
knowledge of the parton densities. The last column of Table~\ref{tab:final_alphas_perPDF} lists the goodness-of-fit
per degree of freedom ($\chi^2$/ndf) of the final single combined result compared to the twelve individual $\alpS(m_{\PZ})$
extractions.

\begin{table}[htbp!]
\topcaption{Strong coupling constant $\alpS(m_{\PZ})$ values extracted per PDF set by combining all the individual results obtained for
each \PWpm and \PZ boson production cross section measurements (Table~\ref{tab:all_alphas}), listed along with their total and individual uncertainties.
The last column tabulates the goodness-of-fit per degree of freedom $\chi^2$/ndf of the final single combined result compared to the
twelve individual $\alpS(m_{\PZ})$ extractions.
\label{tab:final_alphas_perPDF}}
\centering
\cmsTable{
\renewcommand*{\arraystretch}{1.2}
\begin{tabular}{lcccccccc}\hline
PDF & $\alpS(m_{\PZ})$  & $\delta\stat$ & $\delta\lum$ & $\delta\syst$ & $\delta$(PDF) & $\delta$(scale) & $\delta\statt$ & $\chi^2$/ndf \\\hline
CT14       & $0.1163^{+0.0024}_{-0.0031}$ & 0.0007 & 0.0013 & 0.0010 & \,$^{+0.0016}_{-0.0022}$ & 0.0009 & 0.0006 & 13.3/11 \\
HERAPDF2.0 & $0.1072^{+0.0043}_{-0.0040}$ & 0.0012 & 0.0027 & 0.0012 & \,$^{+0.0027}_{-0.0020}$ & 0.0012 & 0.0009 & 14.2/11 \\
MMHT14     & $0.1186 \pm 0.0025$ & 0.0003 & 0.0018 & 0.0009 & 0.0013 & 0.0007 & 0.0002 & 10.2/11 \\
NNPDF3.0   & $0.1147 \pm 0.0023$ & 0.0009 & 0.0008 & 0.0007 & 0.0014 & 0.0006 & 0.0010 & 29.2/11 \\\hline
\end{tabular}
}
\end{table}

\begin{figure}[htbp!]
\centering
\includegraphics[width=0.58\textwidth]{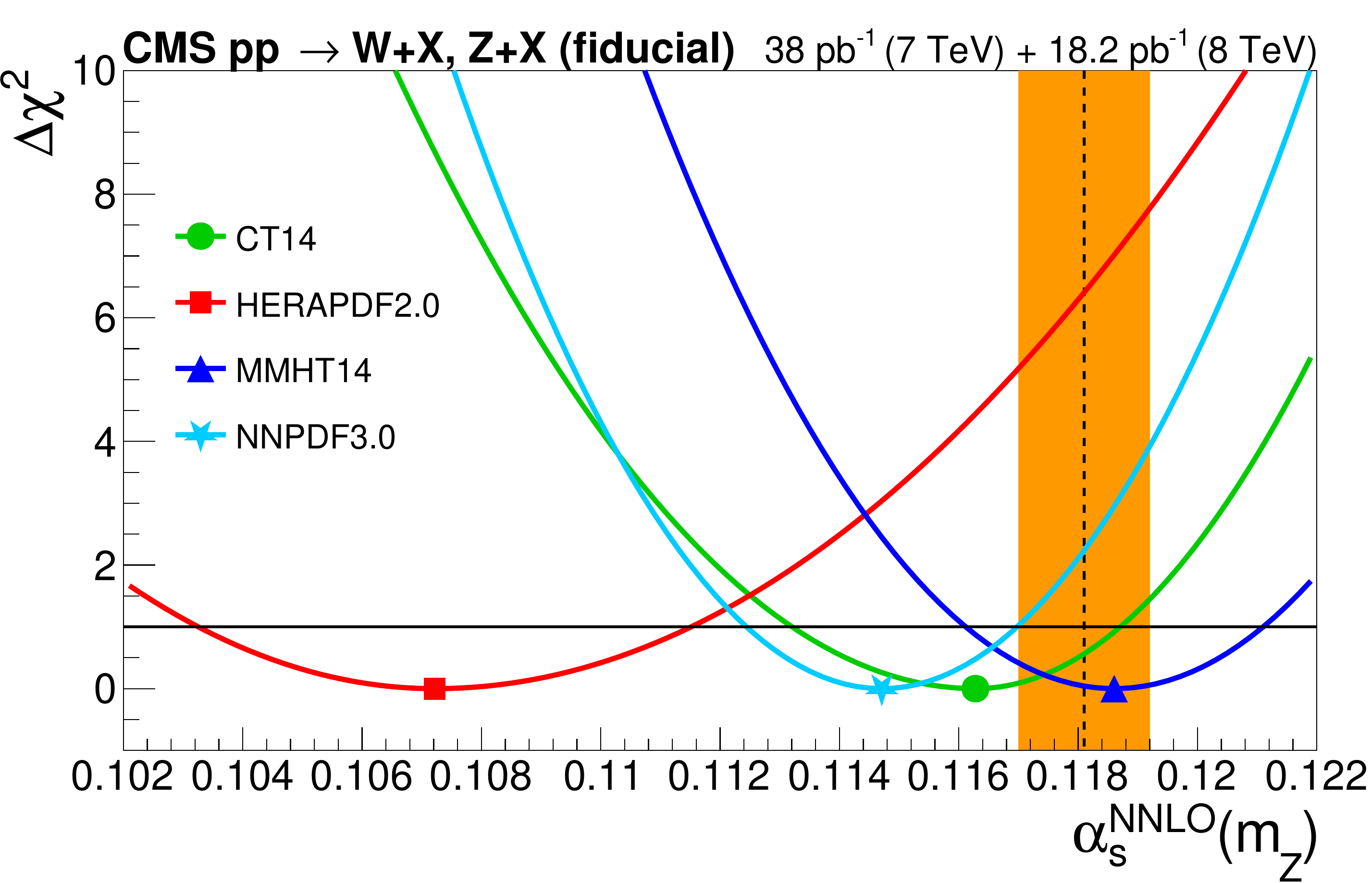}
\includegraphics[width=0.41\textwidth]{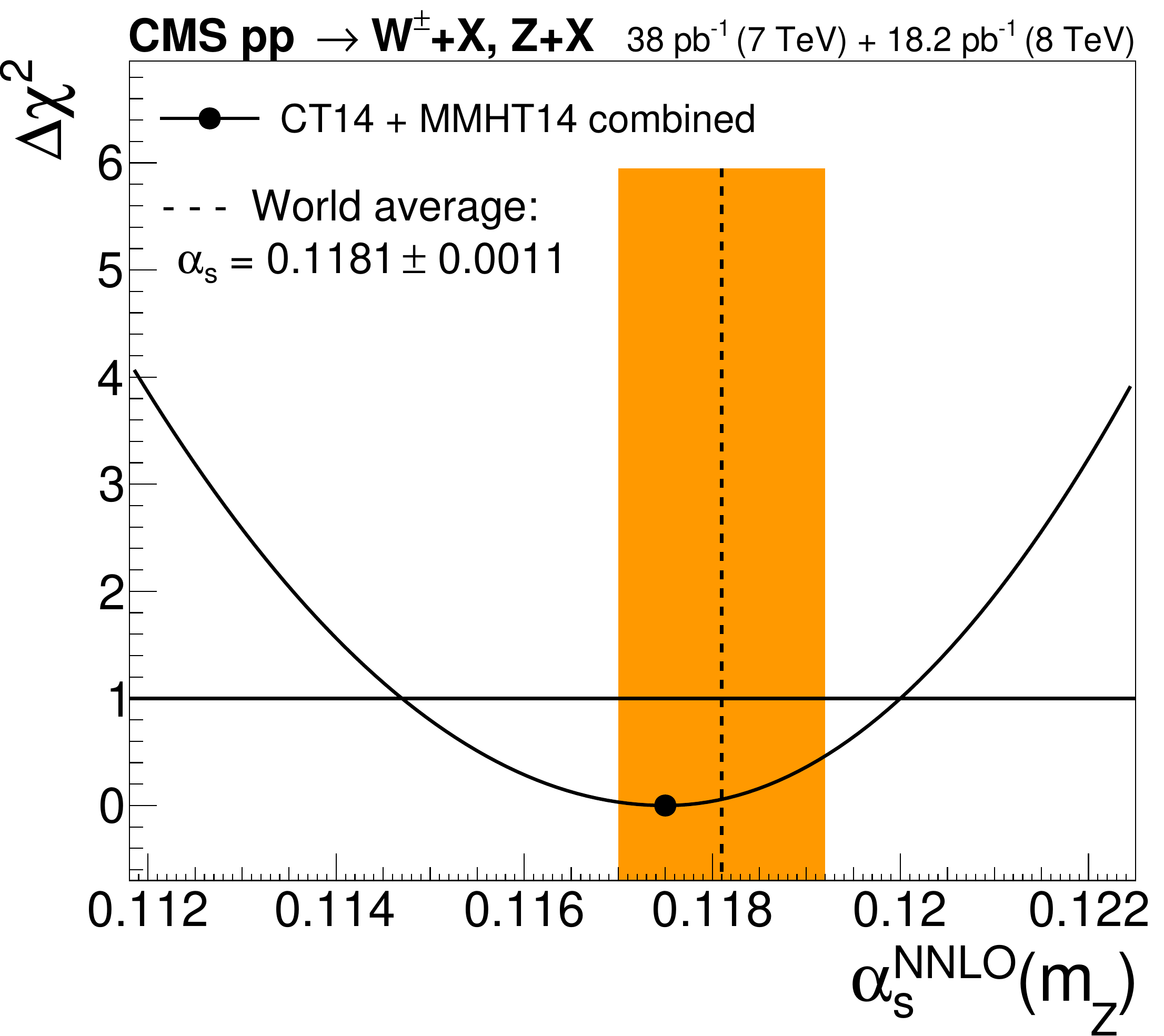}
\caption{Final $\alpS(m_{\PZ})$ values extracted for the CT14, HERAPDF2.0, MMHT14, and NNPDF3.0 PDF sets (left),
and combined $\alpS(m_{\PZ})$ extraction from the CT14 and MMHT14 PDFs (right), compared to the current world average (vertical orange band).
The asymmetric parabolas are constructed to have a minimum at the combined value and are fitted to go through $\Delta\chi^2=1$ (horizontal black lines) at the
one std.\,deviation  uncertainties quoted in Table~\ref{tab:final_alphas_perPDF}.
\label{fig:allestsavgonlyPar}}
\end{figure}

The $\alpS(m_{\PZ})$ results obtained with HERAPDF2.0 and NNPDF3.0 show various differences with respect to those derived
with the CT14 and MMHT14 sets. First, the QCD coupling constant derived with HERAPDF2.0, $\alpS(m_{\PZ}) = 0.1072^{+0.0043}_{-0.0040}$,
is between 1.7 and 2.7 standard deviations smaller than the rest of extractions (Fig.~\ref{fig:allestsavgonlyPar} left).
Although as discussed later, in the context of the cross-checks described in Table~\ref{tab:alphas_sensitivity},
such a disagreement is reduced when symmetrising the HERAPDF2.0 uncertainties to their maximum values.
As discussed before, since the HERAPDF2.0 cross sections for $\alpS(m_{\PZ}) = 0.118$ tend to \textit{overpredict}
the measured \PWpm and \PZ boson production cross sections (Figs.~\ref{fig:fitgraphfirst}--\ref{fig:fitgraphlast}),
a data-theory agreement can only be obtained for a value of $\alpS$ that is \textit{reduced} compared to its default value.
For all global PDF fits extracted with different $\alpS$ values as input, there exists a generic anticorrelation between the values of $\alpS(Q^2)$ and the parton densities evaluated at $(x,Q^2$), particularly, for the gluon and in turn (through perturbative evolution) for the sea quarks. It is thereby important to analyze in more detail the differences between the PDF sets for each flavour. For this purpose, a comparison study of parton densities and parton luminosities has been carried out with \textsc{apfel} v2.7.1~\cite{Carrazza:2014gfa}. This study indicates that the HERAPDF2.0 \cPqu-quark densities (and the overall
quark-antiquark luminosities) are enhanced by about 5\% compared to the rest of PDFs in the $(x,Q^2)$ region of relevance for
EW boson production. This fact increases the weight of the LO contributions to the theoretical \PWpm and \PZ boson
production cross sections,
and thereby pushes down the cross section contributions from higher-order pQCD diagrams that are sensitive to $\alpS(m_{\PZ})$.
The effective result is a comparatively reduced $\alpS(m_{\PZ})$ value.
The level of agreement between the twelve individual and the total $\alpS(m_{\PZ})$ extractions turns out to be good for this PDF set
($\chi^2/\mathrm{ndf}\approx1$ in Table~\ref{tab:final_alphas_perPDF}), because of the relatively wide span of $\alpS$ values
derived and their associated large uncertainties (Fig.~\ref{fig:allests}).
Since HERAPDF2.0 uses DIS data alone, and therefore lacks the extra constraints on the PDFs provided by the LHC data,
we conclude that one would need an updated refit of these parton densities to an extended set of experimental data,
including LHC results, before relying on the QCD coupling constant derived following the procedure described here.

The features of the $\alpS(m_{\PZ})$ results obtained with NNPDF3.0 show the opposite behaviour to those observed for
the HERAPDF2.0 set. The \PWpm and \PZ boson production cross sections computed with this PDF tend to \textit{underpredict} the experimental measurements
(Figs.~\ref{fig:fitgraphfirst}--\ref{fig:fitgraphlast}), and yield an overall bad data-theory agreement ($\chi^2/\mathrm{ndf}\approx3$
in Table~\ref{tab:chi2}), for the baseline $\alpS(m_{\PZ}) = 0.118$ coupling constant. A reproduction of the individual measurements by theory
can thus be achieved only for an $\alpS(m_{\PZ})$ value that is \textit{enlarged} compared to the default value for this PDF set. Thus, many of the individual
$\alpS(m_{\PZ})$ extractions obtained with NNPDF3.0 have values relatively larger than those obtained for other PDFs.
However, the final combined NNPDF3.0 value appears shifted down to $\alpS(m_{\PZ}) = 0.1147\pm 0.0023$
(Table~\ref{tab:final_alphas_perPDF}),
falling outside of the region around $\alpS(m_{\PZ})\approx0.120$ defined by most of the individual estimates (Table~\ref{tab:all_alphas}),
and, consequently, the final level of agreement of the combined and single extractions is poor
($\chi^2/\mathrm{ndf}\approx3$ in Table~\ref{tab:final_alphas_perPDF}).
Such a seemingly counterintuitive behaviour is due to the presence of strong correlations among individual extractions for a fixed value of $\alpS(m_{\PZ})$,
and the fact that the lowest $\alpS(m_{\PZ})$ values derived have smaller uncertainties than the rest, thereby pulling down the final average.
The effect of a combined result lying outside of the range of the input values caused by, \eg\ large underlying nonlinearities among individual estimates, is 
called ``Peelle's pertinent puzzle''~\cite{Neudecker:2014}.
The absence of parametrisation bias in this neural-network PDF results in parton densities that can be less well constrained
(have larger uncertainties) than the rest of PDFs in some regions of phase space. The same \textsc{apfel} v2.7.1 study
mentioned above indicates that the NNPDF3.0 quark-antiquark luminosities tend to be somewhat less precise than those from the other PDF sets
in the relevant $(x,Q^2)$ range for \PWpm and \PZ\ production. A larger span of replicas results in nontrivial correlations among PDF
uncertainties that push the final $\alpS(m_{\PZ})$ value off the individual extractions for each single measurement.
In any case, the latest version released (v3.1) of the NNPDF global fit~\cite{Ball:2017nwa} shows much better agreement
of the theoretical EW boson cross sections with the LHC data for the central $\alpS(m_{\PZ}) = 0.118$ value.
However, this latter NNPDF3.1 set cannot be directly employed to independently extract $\alpS(m_{\PZ})$ from the CMS measurements through the
approach discussed here, as this updated global fit does include already the \textit{absolute} normalisation of a fraction of the \PWpm and \PZ
cross sections used in this work.

The final $\alpS(m_{\PZ})$ extractions are plotted in Fig.~\ref{fig:allestsavgonlyPar} (left)---with (asymmetric, where needed) parabolas
constructed so as to have a minimum at each final central $\alpS(m_{\PZ})$ result and (one standard deviation)
uncertainties matching those listed in Table~\ref{tab:final_alphas_perPDF}---compared with the current world average of $\alpS(m_{\PZ}) = 0.1181 \pm 0.0011$ (orange band).

To analyse the robustness and stability of the final $\alpS(m_{\PZ})$ extractions to the underlying data sets, their uncertainties,
and correlations, we repeat the \textsc{convino} combination varying several ingredients, as explained next. First, $\alpS(m_{\PZ})$ values are extracted
using separately the measurements at $\sqrts = 7$ and 8\TeV alone, as shown in the top rows of Table~\ref{tab:alphas_sensitivity}.
This separation of data sets yields final $\alpS(m_{\PZ})$ values mostly consistent with those derived from the combined ones
listed in Table~\ref{tab:final_alphas_perPDF}, with the largest deviations from the original results being those from
the 7\TeV NNPDF3.0 and 8\TeV HERAPDF2.0 extractions, with 1.0 and 0.85 standard deviations, respectively.
Although the CMS luminosity studies confirm that these uncertainties are fully uncorrelated between 7 and 8\TeV, we have checked the impact of relaxing such an assumption by assuming a 0.5 correlation factor between them. Such a correlation factor results in a change of the final $\alpS(m_{\PZ})$ by at most one-third of the current total uncertainty.
Another cross-check is carried out by symmetrising the PDF uncertainties to their maximum value of the two (this does not apply to NNPDF3.0,
because its uncertainties are symmetric by construction). The corresponding results are given in the left bottom half of
Table~\ref{tab:alphas_sensitivity}. Changing from asymmetric to symmetric PDF uncertainties causes the HERAPDF2.0 combined value to increase
by 1.1 standard deviation, whereas all other PDF results are consistent with the default $\alpS(m_{\PZ})$ extractions. Such a large sensitivity
to changes in the PDF uncertainty confirms the relative lack of robustness of the $\alpS(m_{\PZ})$ values derived for HERAPDF2.0 in our analysis, because the asymmetries of the PDF uncertainties can be significantly affected by random numerical errors.
To further test the sensitivity of the $\alpS(m_{\PZ})$ extraction to the assumptions made on the underlying \PWpm and \PZ
cross section uncertainties and their correlations, the original analysis is repeated by adding an uncorrelated 1\% numerical uncertainty to all theoretical cross sections.
Such a value accounts for possible overlooked small uncorrelated uncertainties, \eg\ coming from the use of different codes for the
theoretical pQCD and/or EW calculations~\cite{Alioli:2016fum}. The impact of such a change is not significant in the final
results, as observed
by comparing the numbers in Table~\ref{tab:final_alphas_perPDF} and those in the bottom-right columns of Table~\ref{tab:alphas_sensitivity}. Further similar tests and
cross-checks have been carried out in a recent $\alpS(m_{\PZ})$
extraction that exploits all the LHC electroweak boson data~\cite{dEnterria:2019aat}
with the same approach used here, which confirm these conclusions.
All these systematic tests indicate that HERAPDF2.0 and NNPDF3.0 have larger variations when changing the ingredients of the combination, but for CT14 and MMHT14 the final $\alpS(m_{\PZ})$ values extracted are reasonably robust
within the quoted uncertainties.

\begin{table}[!htbp]
\topcaption{Sensitivity of the final $\alpS(m_{\PZ})$ extractions per PDF set to various data, uncertainties, and correlation
assumptions. Top rows: Extractions of $\alpS(m_{\PZ})$ using only the 7 and 8\TeV measurements separately.
Bottom rows: Extractions of $\alpS(m_{\PZ})$ when symmetrising the asymmetric PDF uncertainties by taking the maximum of the negative
and positive values (left), and when adding a 1\% uncorrelated uncertainty to all cross sections (right).
\label{tab:alphas_sensitivity}}
\centering
\renewcommand*{\arraystretch}{1.2}
\begin{tabular}{lcc}\hline
PDF & $\alpS(m_{\PZ})$ [7\TeV data] &  $\alpS(m_{\PZ})$ [8\TeV data] \\\hline
CT14       & $0.1158^{+0.0048}_{-0.0052}$ & $0.1174^{+0.0041}_{-0.0037}$ \\
HERAPDF2.0 & $0.1075 \pm 0.0060$ &  $0.1038^{+0.0107}_{-0.0073}$  \\
MMHT14     & $0.1192^{+0.0071}_{-0.0059}$ & $0.1184 \pm 0.0029$ \\
NNPDF3.0   & $0.1123 \pm 0.0032$ &  $0.1148 \pm 0.0031$ \\\hline
PDF & $\alpS(m_{\PZ})$ [symm. PDF uncert.] &  $\alpS(m_{\PZ})$ [$+1\%$ uncorr.\,uncert.]  \\\hline
CT14 & $0.1148 \pm 0.0034$ & $0.1169^{+0.0027}_{-0.0034}$  \\
HERAPDF2.0 & $0.1119 \pm 0.0056$ & $0.1089 \pm 0.0045$ \\
MMHT14 & $0.1185 \pm 0.0028$ &  $0.1186 \pm 0.0026$ \\
NNPDF3.0 & $0.1147 \pm 0.0023$  & $0.1155 \pm 0.0029$ \\\hline
\end{tabular}
\end{table}

Among PDFs, the results obtained using MMHT14 and CT14 feature the largest sensitivity to $\alpS$ variations,
\ie\ they show a larger $k$ slope, Eq.~(\ref{eq:ellipsoids}), compared to those obtained with HERAPDF2.0 and NNPDF3.0
(Figs.~\ref{fig:fitgraphfirst}--\ref{fig:fitgraphlast}).
Since the uncertainty in the $\alpS(m_{\PZ})$ value derived from HERAPDF2.0 is the largest (up to twice as large as some of the other extractions),
because of the absence of constraining LHC input data in this HERA-only PDF fit,
and since the final NNPDF3.0 result
has a larger tension between the combined and individual extractions from each single measurement (Table~\ref{tab:final_alphas_perPDF}),
we consider the values extracted with CT14, $\alpS(m_{\PZ}) = 0.1163^{+0.0024}_{-0.0031}$, and MMHT14, $\alpS(m_{\PZ}) = 0.1186 \pm 0.0025$,
as the most reliable in this analysis.
Providing a single final $\alpS(m_{\PZ})$ value from this study is not obvious because, in general, there is no unique
way to derive a final best estimate of $\alpS$ based on the results obtained from different PDF sets.
An unbiased approach for combining results from different PDFs, in line with the PDG practice~\cite{PDG} as well as with the
procedure employed to produce the PDF4LHC combined PDF set~\cite{Butterworth:2015oua}, is to average them without applying any
further weighting. The same approach was followed also in the similar combination of QCD coupling constant values obtained
from the inclusive $\ttbar$ cross sections~\cite{Klijnsma:2017eqp}. By taking the straight average of the
mean values and of the uncertainties of the individual CT14 and MMHT14 combinations, we obtain a final value of the QCD coupling constant at the
\PZ pole mass, $\alpS(m_{\PZ}) = 0.1175^{+0.0025}_{-0.0028}$, with a total (symmetrised) uncertainty of 2.3\%. Such a result compares
very well with the $\alpS(m_{\PZ}) = 0.1177^{+0.0034}_{-0.0036}$ value, with an uncertainty of $\approx$3\%, extracted from the theoretical
analysis of top pair cross section data~\cite{Klijnsma:2017eqp}.
The right plot of Fig.~\ref{fig:allestsavgonlyPar} shows the $\alpS(m_{\PZ})$ parabola extracted combining the CT14 and MMHT14 results.
This final extraction is fully consistent with the PDG world average (orange band), and has an overall uncertainty similar to that of
other recent determinations at this level of (NNLO) theoretical accuracy, such as those from EW precision
fits~\cite{Haller:2018nnx}, and $\ttbar$ cross sections~\cite{Chatrchyan:2013haa,Klijnsma:2017eqp,Sirunyan:2018goh}.

\section{Summary}
\label{sec:6}

We have used twelve measurements of the inclusive fiducial \PWpm and \PZ production cross sections in proton-proton collisions ($\Pp\Pp$)
at $\sqrts = 7$ and 8\TeV, carried out in the electron and muon decay channels by the CMS experiment, to extract the value of the strong
coupling constant at the \PZ pole mass, $\alpS(m_{\PZ})$.
The procedure is based on a detailed comparison of the measured electroweak boson cross sections to theoretical calculations computed at
next-to-next-to-leading-order accuracy with the CT14, HERAPDF2.0, MMHT14, and NNPDF3.0 parton distribution function (PDF) sets.
The overall data-theory agreement is good within the experimental and theoretical uncertainties.
A $\chi^2$-minimisation procedure has been employed to combine all twelve individual $\alpS$ extractions per PDF set, properly taking into account all individual
sources of experimental and theoretical uncertainties, and their correlations. The following combined values are extracted for the four different PDFs:
$\alpS(m_{\PZ}) = 0.1163^{+0.0024}_{-0.0031}$ (CT14), $0.1072^{+0.0043}_{-0.0040}$ (HERAPDF2.0), $0.1186\pm0.0025$ (MMHT14), and
$0.1147\pm 0.0023$ (NNPDF3.0). The largest propagated uncertainties are associated with the experimental integrated luminosity
and theoretical intra-PDF uncertainties. Among the four extractions, the cross section calculated with the CT14 and MMHT14 sets
appear as the most sensitive to the underlying $\alpS$ value and, at the same time, the derived $\alpS(m_{\PZ})$ values are the most
robust and stable with respect to variations in the data and theoretical cross sections, their uncertainties, and correlations.
The result derived combining the CT14 and MMHT14 extractions, $\alpS(m_{\PZ}) = 0.1175^{+0.0025}_{-0.0028}$,
has a 2.3\% uncertainty that is comparable to that previously obtained in a similar analysis
of the inclusive $\ttbar$ cross sections in $\Pp\Pp$ collisions at the LHC. This extracted value is fully compatible with the
current $\alpS(m_{\PZ})$ world average.

\begin{acknowledgments}

We congratulate our colleagues in the CERN accelerator departments for the excellent performance of the LHC and thank the technical and administrative staffs at CERN and at other CMS institutes for their contributions to the success of the CMS effort. In addition, we gratefully acknowledge the computing centres and personnel of the Worldwide LHC Computing Grid for delivering so effectively the computing infrastructure essential to our analyses. Finally, we acknowledge the enduring support for the construction and operation of the LHC and the CMS detector provided by the following funding agencies: BMBWF and FWF (Austria); FNRS and FWO (Belgium); CNPq, CAPES, FAPERJ, FAPERGS, and FAPESP (Brazil); MES (Bulgaria); CERN; CAS, MoST, and NSFC (China); COLCIENCIAS (Colombia); MSES and CSF (Croatia); RPF (Cyprus); SENESCYT (Ecuador); MoER, ERC IUT, PUT and ERDF (Estonia); Academy of Finland, MEC, and HIP (Finland); CEA and CNRS/IN2P3 (France); BMBF, DFG, and HGF (Germany); GSRT (Greece); NKFIA (Hungary); DAE and DST (India); IPM (Iran); SFI (Ireland); INFN (Italy); MSIP and NRF (Republic of Korea); MES (Latvia); LAS (Lithuania); MOE and UM (Malaysia); BUAP, CINVESTAV, CONACYT, LNS, SEP, and UASLP-FAI (Mexico); MOS (Montenegro); MBIE (New Zealand); PAEC (Pakistan); MSHE and NSC (Poland); FCT (Portugal); JINR (Dubna); MON, RosAtom, RAS, RFBR, and NRC KI (Russia); MESTD (Serbia); SEIDI, CPAN, PCTI, and FEDER (Spain); MOSTR (Sri Lanka); Swiss Funding Agencies (Switzerland); MST (Taipei); ThEPCenter, IPST, STAR, and NSTDA (Thailand); TUBITAK and TAEK (Turkey); NASU (Ukraine); STFC (United Kingdom); DOE and NSF (USA).

\hyphenation{Rachada-pisek} Individuals have received support from the Marie-Curie programme and the European Research Council and Horizon 2020 Grant, contract Nos.\ 675440, 752730, and 765710 (European Union); the Leventis Foundation; the A.P.\ Sloan Foundation; the Alexander von Humboldt Foundation; the Belgian Federal Science Policy Office; the Fonds pour la Formation \`a la Recherche dans l'Industrie et dans l'Agriculture (FRIA-Belgium); the Agentschap voor Innovatie door Wetenschap en Technologie (IWT-Belgium); the F.R.S.-FNRS and FWO (Belgium) under the ``Excellence of Science -- EOS" -- be.h project n.\ 30820817; the Beijing Municipal Science \& Technology Commission, No. Z181100004218003; the Ministry of Education, Youth and Sports (MEYS) of the Czech Republic; the Deutsche Forschungsgemeinschaft (DFG) under Germany's Excellence Strategy -- EXC 2121 ``Quantum Universe" -- 390833306; the Lend\"ulet (``Momentum") Programme and the J\'anos Bolyai Research Scholarship of the Hungarian Academy of Sciences, the New National Excellence Program \'UNKP, the NKFIA research grants 123842, 123959, 124845, 124850, 125105, 128713, 128786, and 129058 (Hungary); the Council of Science and Industrial Research, India; the HOMING PLUS programme of the Foundation for Polish Science, cofinanced from European Union, Regional Development Fund, the Mobility Plus programme of the Ministry of Science and Higher Education, the National Science Center (Poland), contracts Harmonia 2014/14/M/ST2/00428, Opus 2014/13/B/ST2/02543, 2014/15/B/ST2/03998, and 2015/19/B/ST2/02861, Sonata-bis 2012/07/E/ST2/01406; the National Priorities Research Program by Qatar National Research Fund; the Ministry of Science and Education, grant no. 3.2989.2017 (Russia); the Programa Estatal de Fomento de la Investigaci{\'o}n Cient{\'i}fica y T{\'e}cnica de Excelencia Mar\'{\i}a de Maeztu, grant MDM-2015-0509 and the Programa Severo Ochoa del Principado de Asturias; the Thalis and Aristeia programmes cofinanced by EU-ESF and the Greek NSRF; the Rachadapisek Sompot Fund for Postdoctoral Fellowship, Chulalongkorn University and the Chulalongkorn Academic into Its 2nd Century Project Advancement Project (Thailand); the Nvidia Corporation; the Welch Foundation, contract C-1845; and the Weston Havens Foundation (USA).
\end{acknowledgments}

\bibliography{auto_generated}

\clearpage

\appendix

\section{Correlation matrices of the experimental measurements}

Relevant correlation matrices of the experimental systematic uncertainties in the measurements of \PWpm and \PZ boson production cross
sections, and in their associated extractions of $\alpS(m_{\PZ})$, are listed in Tables~\ref{tab:expcorrfirst}--\ref{tab:expcorrlast}
and~\ref{tab:exptotalcorr}--\ref{tab:totalcorr1}, respectively.

\begin{table}[!hbt]
\topcaption{Experimental systematic uncertainties: Lepton reconstruction and identification correlation matrix. \label{tab:expcorrfirst}}
\centering
\begin{tabular}{cc|cccccc | cccccc}\hline
&& \multicolumn{6}{c}{7\TeV} & \multicolumn{6}{c}{8\TeV} \\
&& $\PW^{+}_{\Pe}$ & $\PW^{-}_{\Pe}$ & $\PZ_{\Pe}$ & $\PW^{+}_{\PGm}$ & $\PW^{-}_{\PGm}$ & $\PZ_{\PGm}$ & $\PW^{+}_{\Pe}$ & $\PW^{-}_{\Pe}$ & $\PZ_{\Pe}$ & $\PW^{+}_{\PGm}$ & $\PW^{-}_{\PGm}$ & $\PZ_{\PGm}$ \\\hline
& $\PW^{+}_{\Pe}$& 1 & 1 & 1 & 0 & 0 & 0 & 1 & 1 & 1 & 0 & 0 & 0 \\
& $\PW^{-}_{\Pe}$& 1 & 1 & 1 & 0 & 0 & 0 & 1 & 1 & 1 & 0 & 0 & 0 \\
7& $\PZ_{\Pe}$& 1 & 1 & 1 & 0 & 0 & 0 & 1 & 1 & 1 & 0 & 0 & 0 \\
TeV& $\PW^{+}_{\PGm}$& 0 & 0 & 0 & 1 & 1 & 1 & 0 & 0 & 0 & 1 & 1 & 1 \\
& $\PW^{-}_{\PGm}$& 0 & 0 & 0 & 1 & 1 & 1 & 0 & 0 & 0 & 1 & 1 & 1 \\
& $\PZ_{\PGm}$& 0 & 0 & 0 & 1 & 1 & 1 & 0 & 0 & 0 & 1 & 1 & 1 \\ \hline
& $\PW^{+}_{\Pe}$& 1 & 1 & 1 & 0 & 0 & 0 & 1 & 1 & 1 & 0 & 0 & 0 \\
& $\PW^{-}_{\Pe}$& 1 & 1 & 1 & 0 & 0 & 0 & 1 & 1 & 1 & 0 & 0 & 0 \\
8& $\PZ_{\Pe}$& 1 & 1 & 1 & 0 & 0 & 0 & 1 & 1 & 1 & 0 & 0 & 0 \\
TeV& $\PW^{+}_{\PGm}$& 0 & 0 & 0 & 1 & 1 & 1 & 0 & 0 & 0 & 1 & 1 & 1 \\
& $\PW^{-}_{\PGm}$& 0 & 0 & 0 & 1 & 1 & 1 & 0 & 0 & 0 & 1 & 1 & 1 \\
& $\PZ_{\PGm}$& 0 & 0 & 0 & 1 & 1 & 1 & 0 & 0 & 0 & 1 & 1 & 1 \\ \hline
\end{tabular}
\end{table}

\begin{table}[!hbt]
\topcaption{Experimental systematic uncertainties: Muon trigger inefficiency correlation matrix~\cite{Khachatryan:2016bia}.}
\centering
\begin{tabular}{cc|cccccc | cccccc}\hline
&& \multicolumn{6}{c}{7\TeV} & \multicolumn{6}{c}{8\TeV} \\
&& $\PW^{+}_{\Pe}$ & $\PW^{-}_{\Pe}$ & $\PZ_{\Pe}$ & $\PW^{+}_{\PGm}$ & $\PW^{-}_{\PGm}$ & $\PZ_{\PGm}$ & $\PW^{+}_{\Pe}$ & $\PW^{-}_{\Pe}$ & $\PZ_{\Pe}$ & $\PW^{+}_{\PGm}$ & $\PW^{-}_{\PGm}$ & $\PZ_{\PGm}$ \\\hline
& $\PW^{+}_{\Pe}$& 0 & 0 & 0 & 0 & 0 & 0 & 0 & 0 & 0 & 0 & 0 & 0 \\
& $\PW^{-}_{\Pe}$& 0 & 0 & 0 & 0 & 0 & 0 & 0 & 0 & 0 & 0 & 0 & 0 \\
7& $\PZ_{\Pe}$& 0 & 0 & 0 & 0 & 0 & 0 & 0 & 0 & 0 & 0 & 0 & 0 \\
TeV& $\PW^{+}_{\PGm}$& 0 & 0 & 0 & 1 & 1 & 1 & 0 & 0 & 0 & 1 & 1 & 1 \\
& $\PW^{-}_{\PGm}$& 0 & 0 & 0 & 1 & 1 & 1 & 0 & 0 & 0 & 1 & 1 & 1 \\
& $\PZ_{\PGm}$& 0 & 0 & 0 & 1 & 1 & 1 & 0 & 0 & 0 & 1 & 1 & 1 \\ \hline
& $\PW^{+}_{\Pe}$& 0 & 0 & 0 & 0 & 0 & 0 & 0 & 0 & 0 & 0 & 0 & 0 \\
& $\PW^{-}_{\Pe}$& 0 & 0 & 0 & 0 & 0 & 0 & 0 & 0 & 0 & 0 & 0 & 0 \\
8& $\PZ_{\Pe}$& 0 & 0 & 0 & 0 & 0 & 0 & 0 & 0 & 0 & 0 & 0 & 0 \\
TeV& $\PW^{+}_{\PGm}$& 0 & 0 & 0 & 1 & 1 & 1 & 0 & 0 & 0 & 1 & 1 & 1 \\
& $\PW^{-}_{\PGm}$& 0 & 0 & 0 & 1 & 1 & 1 & 0 & 0 & 0 & 1 & 1 & 1 \\
& $\PZ_{\PGm}$& 0 & 0 & 0 & 1 & 1 & 1 & 0 & 0 & 0 & 1 & 1 & 1 \\ \hline
\end{tabular}
\end{table}

\begin{table}[!hbt]
\topcaption{Experimental systematic uncertainties: Energy and momentum scale and resolution correlation matrix.}
\centering
\begin{tabular}{cc|cccccc | cccccc}\hline
&& \multicolumn{6}{c}{7\TeV} & \multicolumn{6}{c}{8\TeV} \\
&& $\PW^{+}_{\Pe}$ & $\PW^{-}_{\Pe}$ & $\PZ_{\Pe}$ & $\PW^{+}_{\PGm}$ & $\PW^{-}_{\PGm}$ & $\PZ_{\PGm}$ & $\PW^{+}_{\Pe}$ & $\PW^{-}_{\Pe}$ & $\PZ_{\Pe}$ & $\PW^{+}_{\PGm}$ & $\PW^{-}_{\PGm}$ & $\PZ_{\PGm}$ \\\hline
& $\PW^{+}_{\Pe}$& 1 & 1 & 1 & 0 & 0 & 0 & 1 & 1 & 1 & 0 & 0 & 0 \\
& $\PW^{-}_{\Pe}$& 1 & 1 & 1 & 0 & 0 & 0 & 1 & 1 & 1 & 0 & 0 & 0 \\
7& $\PZ_{\Pe}$& 1 & 1 & 1 & 0 & 0 & 0 & 1 & 1 & 1 & 0 & 0 & 0 \\
TeV& $\PW^{+}_{\PGm}$& 0 & 0 & 0 & 1 & 1 & 1 & 0 & 0 & 0 & 1 & 1 & 1 \\
& $\PW^{-}_{\PGm}$& 0 & 0 & 0 & 1 & 1 & 1 & 0 & 0 & 0 & 1 & 1 & 1 \\
& $\PZ_{\PGm}$& 0 & 0 & 0 & 1 & 1 & 1 & 0 & 0 & 0 & 1 & 1 & 1 \\ \hline
& $\PW^{+}_{\Pe}$& 1 & 1 & 1 & 0 & 0 & 0 & 1 & 1 & 1 & 0 & 0 & 0 \\
& $\PW^{-}_{\Pe}$& 1 & 1 & 1 & 0 & 0 & 0 & 1 & 1 & 1 & 0 & 0 & 0 \\
8& $\PZ_{\Pe}$& 1 & 1 & 1 & 0 & 0 & 0 & 1 & 1 & 1 & 0 & 0 & 0 \\
TeV& $\PW^{+}_{\PGm}$& 0 & 0 & 0 & 1 & 1 & 1 & 0 & 0 & 0 & 1 & 1 & 1 \\
& $\PW^{-}_{\PGm}$& 0 & 0 & 0 & 1 & 1 & 1 & 0 & 0 & 0 & 1 & 1 & 1 \\
& $\PZ_{\PGm}$& 0 & 0 & 0 & 1 & 1 & 1 & 0 & 0 & 0 & 1 & 1 & 1 \\ \hline
\end{tabular}
\end{table}

\begin{table}[!hbt]
\topcaption{Experimental systematic uncertainties: Missing $\pt$ scale and resolution correlation matrix.}
\centering
\begin{tabular}{cc|cccccc | cccccc}\hline
&& \multicolumn{6}{c}{7\TeV} & \multicolumn{6}{c}{8\TeV} \\
&& $\PW^{+}_{\Pe}$ & $\PW^{-}_{\Pe}$ & $\PZ_{\Pe}$ & $\PW^{+}_{\PGm}$ & $\PW^{-}_{\PGm}$ & $\PZ_{\PGm}$ & $\PW^{+}_{\Pe}$ & $\PW^{-}_{\Pe}$ & $\PZ_{\Pe}$ & $\PW^{+}_{\PGm}$ & $\PW^{-}_{\PGm}$ & $\PZ_{\PGm}$ \\\hline
& $\PW^{+}_{\Pe}$& 1 & 1 & 0 & 1 & 1 & 0 & 1 & 1 & 0 & 1 & 1 & 0 \\
& $\PW^{-}_{\Pe}$& 1 & 1 & 0 & 1 & 1 & 0 & 1 & 1 & 0 & 1 & 1 & 0 \\
7& $\PZ_{\Pe}$& 0 & 0 & 1 & 0 & 0 & 0 & 0 & 0 & 1 & 0 & 0 & 0 \\
TeV& $\PW^{+}_{\PGm}$& 1 & 1 & 0 & 1 & 1 & 0 & 1 & 1 & 0 & 1 & 1 & 0 \\
& $\PW^{-}_{\PGm}$& 1 & 1 & 0 & 1 & 1 & 0 & 1 & 1 & 0 & 1 & 1 & 0 \\
& $\PZ_{\PGm}$& 0 & 0 & 0 & 0 & 0 & 1 & 0 & 0 & 1 & 0 & 0 & 1 \\ \hline
& $\PW^{+}_{\Pe}$& 1 & 1 & 0 & 1 & 1 & 0 & 1 & 1 & 0 & 1 & 1 & 0 \\
& $\PW^{-}_{\Pe}$& 1 & 1 & 0 & 1 & 1 & 0 & 1 & 1 & 0 & 1 & 1 & 0 \\
8& $\PZ_{\Pe}$& 0 & 0 & 1 & 0 & 0 & 0 & 0 & 0 & 1 & 0 & 0 & 0 \\
TeV& $\PW^{+}_{\PGm}$& 1 & 1 & 0 & 1 & 1 & 0 & 1 & 1 & 0 & 1 & 1 & 0 \\
& $\PW^{-}_{\PGm}$& 1 & 1 & 0 & 1 & 1 & 0 & 1 & 1 & 0 & 1 & 1 & 0 \\
& $\PZ_{\PGm}$& 0 & 0 & 0 & 0 & 0 & 1 & 0 & 0 & 0 & 0 & 0 & 1 \\ \hline
\end{tabular}
\end{table}

\begin{table}[!hbt]
\topcaption{Experimental systematic uncertainties: Background subtraction and modelling correlation matrix.\label{tab:expcorrlast}}
\centering
\begin{tabular}{cc|cccccc | cccccc}\hline
&& \multicolumn{6}{c}{7\TeV} & \multicolumn{6}{c}{8\TeV} \\
&& $\PW^{+}_{\Pe}$ & $\PW^{-}_{\Pe}$ & $\PZ_{\Pe}$ & $\PW^{+}_{\PGm}$ & $\PW^{-}_{\PGm}$ & $\PZ_{\PGm}$ & $\PW^{+}_{\Pe}$ & $\PW^{-}_{\Pe}$ & $\PZ_{\Pe}$ & $\PW^{+}_{\PGm}$ & $\PW^{-}_{\PGm}$ & $\PZ_{\PGm}$ \\\hline
& $\PW^{+}_{\Pe}$& 1 & 1 & 0 & 1 & 1 & 0 & 1 & 1 & 0 & 1 & 1 & 0 \\
& $\PW^{-}_{\Pe}$& 1 & 1 & 0 & 1 & 1 & 0 & 1 & 1 & 0 & 1 & 1 & 0 \\
7& $\PZ_{\Pe}$& 0 & 0 & 1 & 0 & 0 & 1 & 0 & 0 & 1 & 0 & 0 & 1 \\
TeV& $\PW^{+}_{\PGm}$& 1 & 1 & 0 & 1 & 1 & 0 & 1 & 1 & 0 & 1 & 1 & 0 \\
& $\PW^{-}_{\PGm}$& 1 & 1 & 0 & 1 & 1 & 0 & 1 & 1 & 0 & 1 & 1 & 0 \\
& $\PZ_{\PGm}$& 0 & 0 & 1 & 0 & 0 & 1 & 0 & 0 & 1 & 0 & 0 & 1 \\ \hline
& $\PW^{+}_{\Pe}$& 1 & 1 & 0 & 1 & 1 & 0 & 1 & 1 & 0 & 1 & 1 & 0 \\
& $\PW^{-}_{\Pe}$& 1 & 1 & 0 & 1 & 1 & 0 & 1 & 1 & 0 & 1 & 1 & 0 \\
8& $\PZ_{\Pe}$& 0 & 0 & 1 & 0 & 0 & 1 & 0 & 0 & 1 & 0 & 0 & 1 \\
TeV& $\PW^{+}_{\PGm}$& 1 & 1 & 0 & 1 & 1 & 0 & 1 & 1 & 0 & 1 & 1 & 0 \\
& $\PW^{-}_{\PGm}$& 1 & 1 & 0 & 1 & 1 & 0 & 1 & 1 & 0 & 1 & 1 & 0 \\
& $\PZ_{\PGm}$& 0 & 0 & 1 & 0 & 0 & 1 & 0 & 0 & 1 & 0 & 0 & 1 \\ \hline
\end{tabular}
\end{table}

\begin{table}[!hbt]
\topcaption{Total experimental systematic uncertainties correlations among $\alpS(m_{\PZ})$ values extracted
from all individual measurements at 7 and 8\TeV~\cite{CMS:2011aa,Chatrchyan:2014mua}
of the \PWpm and \PZ boson production cross sections.
\label{tab:exptotalcorr}}
\centering
\cmsTable{
\begin{tabular}{cc|cccccc|cccccc}\hline
&& \multicolumn{6}{c}{7\TeV} & \multicolumn{6}{c}{8\TeV} \\
&& $\PW^{+}_{\Pe}$ & $\PW^{-}_{\Pe}$ & $\PZ_{\Pe}$ & $\PW^{+}_{\PGm}$ & $\PW^{-}_{\PGm}$ & $\PZ_{\PGm}$ & $\PW^{+}_{\Pe}$ & $\PW^{-}_{\Pe}$ & $\PZ_{\Pe}$ & $\PW^{+}_{\PGm}$ & $\PW^{-}_{\PGm}$ & $\PZ_{\PGm}$ \\\hline
& $\PW^{+}_{\Pe}$& 1.000 & 0.992 & 0.932 & 0.097 & 0.108 & 0.000 & 0.974 & 0.993 & 0.907 & 0.109 & 0.102 & 0.000 \\
& $\PW^{-}_{\Pe}$& 0.992 & 1.000 & 0.892 & 0.133 & 0.152 & 0.000 & 0.945 & 0.975 & 0.865 & 0.124 & 0.108 & 0.000 \\
7& $\PZ_{\Pe}$& 0.932 & 0.892 & 1.000 & 0.000 & 0.000 & 0.032 & 0.954 & 0.936 & 0.996 & 0.000 & 0.000 & 0.026 \\
TeV& $\PW^{+}_{\PGm}$& 0.097 & 0.133 & 0.000 & 1.000 & 0.996 & 0.414 & 0.072 & 0.084 & 0.000 & 0.850 & 0.821 & 0.743 \\
& $\PW^{-}_{\PGm}$& 0.108 & 0.152 & 0.000 & 0.996 & 1.000 & 0.423 & 0.074 & 0.090 & 0.000 & 0.842 & 0.809 & 0.712 \\
& $\PZ_{\PGm}$& 0.000 & 0.000 & 0.032 & 0.414 & 0.423 & 1.000 & 0.000 & 0.000 & 0.059 & 0.133 & 0.145 & 0.143 \\\hline
& $\PW^{+}_{\Pe}$& 0.974 & 0.945 & 0.954 & 0.072 & 0.074 & 0.000 & 1.000 & 0.991 & 0.941 & 0.127 & 0.132 & 0.000 \\
& $\PW^{-}_{\Pe}$& 0.993 & 0.975 & 0.936 & 0.084 & 0.090 & 0.000 & 0.991 & 1.000 & 0.915 & 0.129 & 0.130 & 0.000 \\
8& $\PZ_{\Pe}$& 0.907 & 0.865 & 0.996 & 0.000 & 0.000 & 0.059 & 0.941 & 0.915 & 1.000 & 0.000 & 0.000 & 0.048 \\
TeV& $\PW^{+}_{\PGm}$& 0.109 & 0.124 & 0.000 & 0.850 & 0.842 & 0.133 & 0.127 & 0.129 & 0.000 & 1.000 & 0.996 & 0.800 \\
& $\PW^{-}_{\PGm}$& 0.102 & 0.108 & 0.000 & 0.821 & 0.809 & 0.145 & 0.132 & 0.130 & 0.000 & 0.996 & 1.000 & 0.785 \\
& $\PZ_{\PGm}$& 0.000 & 0.000 & 0.026 & 0.743 & 0.712 & 0.143 & 0.000 & 0.000 & 0.048 & 0.800 & 0.785 & 1.000 \\\hline
\end{tabular}
}
\end{table}

\begin{table}[!hbt]
\topcaption{Total experimental and theoretical correlations between all the $\alpS(m_{\PZ})$ values extracted from all measurements
at 7 and 8\TeV~\cite{CMS:2011aa,Chatrchyan:2014mua} of \PWpm and \PZ boson production cross sections,
obtained using the NNPDF3.0 PDF set.\label{tab:totalcorr1}}
\centering
\cmsTable{
\begin{tabular}{cc|cccccc|cccccc}\hline
&& \multicolumn{6}{c}{7\TeV} & \multicolumn{6}{c}{8\TeV} \\
&& $\PW^{+}_{\Pe}$ & $\PW^{-}_{\Pe}$ & $\PZ_{\Pe}$ & $\PW^{+}_{\PGm}$ & $\PW^{-}_{\PGm}$ & $\PZ_{\PGm}$ & $\PW^{+}_{\Pe}$ & $\PW^{-}_{\Pe}$ & $\PZ_{\Pe}$ & $\PW^{+}_{\PGm}$ & $\PW^{-}_{\PGm}$ & $\PZ_{\PGm}$ \\\hline
& $\PW^{+}_{\Pe}$	& 1.000 & 0.976 & 0.933 & 0.857 & 0.856 & 0.818 & 0.426 & 0.444 & 0.360 & 0.243 & 0.243 & 0.161 \\
& $\PW^{-}_{\Pe}$	& 0.976 & 1.000 & 0.938 & 0.871 & 0.880 & 0.835 & 0.417 & 0.435 & 0.353 & 0.248 & 0.254 & 0.168 \\
7& $\PZ_{\Pe}$	& 0.933 & 0.938 & 1.000 & 0.797 & 0.817 & 0.803 & 0.443 & 0.430 & 0.392 & 0.210 & 0.210 & 0.167 \\
TeV& $\PW^{+}_{\PGm}$	& 0.857 & 0.871 & 0.797 & 1.000 & 0.967 & 0.893 & 0.200 & 0.236 & 0.189 & 0.343 & 0.322 & 0.289 \\
& $\PW^{-}_{\PGm}$	& 0.856 & 0.880 & 0.817 & 0.967 & 1.000 & 0.908 & 0.222 & 0.239 & 0.198 & 0.339 & 0.328 & 0.292 \\
& $\PZ_{\PGm}$	& 0.818 & 0.835 & 0.803 & 0.893 & 0.908 & 1.000 & 0.177 & 0.201 & 0.195 & 0.242 & 0.246 & 0.199 \\\hline
& $\PW^{+}_{\Pe}$	& 0.426 & 0.417 & 0.443 & 0.200 & 0.222 & 0.177 & 1.000 & 0.940 & 0.794 & 0.664 & 0.685 & 0.467 \\
& $\PW^{-}_{\Pe}$	& 0.444 & 0.435 & 0.430 & 0.236 & 0.239 & 0.201 & 0.940 & 1.000 & 0.816 & 0.730 & 0.761 & 0.520 \\
8& $\PZ_{\Pe}$	& 0.360 & 0.353 & 0.392 & 0.189 & 0.198 & 0.195 & 0.794 & 0.816 & 1.000 & 0.612 & 0.643 & 0.472 \\
TeV& $\PW^{+}_{\PGm}$	& 0.243 & 0.248 & 0.210 & 0.343 & 0.339 & 0.242 & 0.664 & 0.730 & 0.612 & 1.000 & 0.934 & 0.740 \\
& $\PW^{-}_{\PGm}$	& 0.243 & 0.254 & 0.210 & 0.322 & 0.328 & 0.246 & 0.685 & 0.761 & 0.643 & 0.934 & 1.000 & 0.729 \\
& $\PZ_{\PGm}$	& 0.161 & 0.168 & 0.167 & 0.289 & 0.292 & 0.199 & 0.467 & 0.520 & 0.472 & 0.740 & 0.729 & 1.000 \\\hline
\end{tabular}
}
\end{table}
\cleardoublepage \section{The CMS Collaboration \label{app:collab}}\begin{sloppypar}\hyphenpenalty=5000\widowpenalty=500\clubpenalty=5000\vskip\cmsinstskip
\textbf{Yerevan Physics Institute, Yerevan, Armenia}\\*[0pt]
A.M.~Sirunyan$^{\textrm{\dag}}$, A.~Tumasyan
\vskip\cmsinstskip
\textbf{Institut f\"{u}r Hochenergiephysik, Wien, Austria}\\*[0pt]
W.~Adam, F.~Ambrogi, T.~Bergauer, M.~Dragicevic, J.~Er\"{o}, A.~Escalante~Del~Valle, M.~Flechl, R.~Fr\"{u}hwirth\cmsAuthorMark{1}, M.~Jeitler\cmsAuthorMark{1}, N.~Krammer, I.~Kr\"{a}tschmer, D.~Liko, T.~Madlener, I.~Mikulec, N.~Rad, J.~Schieck\cmsAuthorMark{1}, R.~Sch\"{o}fbeck, M.~Spanring, W.~Waltenberger, C.-E.~Wulz\cmsAuthorMark{1}, M.~Zarucki
% ...
\vskip\cmsinstskip
\dag: Deceased\\
1:  Also at Vienna University of Technology, Vienna, Austria\\
% ...
\end{sloppypar}
\end{document}